\documentclass{article}
\usepackage{iclr2027_conference,times}

\iclrfinalcopy

\usepackage{amsmath,amsfonts,bm}

\def\eqref#1{equation~\ref{#1}}

\def\1{\bm{1}}

\DeclareMathAlphabet{\mathsfit}{\encodingdefault}{\sfdefault}{m}{sl}
\SetMathAlphabet{\mathsfit}{bold}{\encodingdefault}{\sfdefault}{bx}{n}

\usepackage[utf8]{inputenc}
\usepackage[T1]{fontenc}
\usepackage{amsmath,amssymb,mathtools}
\usepackage{textgreek}
\usepackage{graphicx}
\usepackage{booktabs}
\usepackage{multirow}
\usepackage{array}
\usepackage{tabularx}
\usepackage[table]{xcolor}
\usepackage{adjustbox}
\usepackage{microtype}
\usepackage{xcolor}
\usepackage{url}
\usepackage{enumitem}
\usepackage{siunitx}
\usepackage{pifont}
\usepackage{subcaption}
\usepackage{makecell}
\usepackage{bm}
\usepackage{xspace}
\usepackage{threeparttable}

\usepackage[colorlinks=true,citecolor=brown,urlcolor=gray]{hyperref}

\setlist[itemize]{leftmargin=*,nosep}
\setlist[enumerate]{leftmargin=*,nosep}

\newif\ifdraftpaper
\draftpapertrue

\definecolor{todoRed}{RGB}{176,30,45}
\definecolor{placeholderBlue}{RGB}{45,91,150}
\definecolor{lightGray}{RGB}{246,246,246}
\definecolor{OursRow}{HTML}{EAF2FB}

\newcommand{\model}{\textsc{TransDubber}\xspace}
\newcommand{\uniss}{UniSS\xspace}

\newcommand{\durat}[1]{\mathrm{Dur@}#1}

\newcommand{\placeholderbox}[2][0.22\textheight]{%
  \begingroup
  \setlength{\fboxsep}{8pt}%
  \noindent\fcolorbox{placeholderBlue}{lightGray}{%
    \begin{minipage}[c][#1][c]{%
      \dimexpr\linewidth-2\fboxsep-2\fboxrule\relax}
      \centering\small
      #2
    \end{minipage}%
  }%
  \endgroup
}

\newcommand{\maybegraphic}[3][0.22\textheight]{%
  \IfFileExists{#2}{%
    \includegraphics[width=\linewidth]{#2}%
  }{%
    \placeholderbox[#1]{%
      \textbf{Planned figure}\par
      \medskip
      #3\par
      \medskip
      \nolinkurl{#2}
    }%
  }%
}

\newcolumntype{Y}{>{\centering\arraybackslash}X}
\newcolumntype{L}[1]{>{\raggedright\arraybackslash}p{#1}}
\newcolumntype{C}[1]{>{\centering\arraybackslash}p{#1}}

\newcommand{\papertitle}{Learning Timeline-Preserving Speech-to-Speech Translation}
\title{\papertitle}
\author{Zhisheng Zheng, David Harwath\\
\normalfont The University of Texas at Austin}

\begin{document}
\maketitle
\lhead{\small\papertitle}
\thispagestyle{plain}

\begin{abstract}

Speech-to-speech translation (S2ST) increasingly preserves \textit{what} is said and \textit{how} it sounds, but not necessarily \textit{when} speech and silence occur.
For dubbing and other temporally grounded applications, matching total duration is not enough: speech and pauses must also occur at corresponding moments.
We formulate \emph{timeline-preserving S2ST}, which aims to preserve the source endpoint and salient internal pauses on a shared absolute timeline.
We present \model{}, an end-to-end autoregressive system that learns this temporal structure directly from aligned target waveforms, without inference-time frame-level timing control or post-hoc temporal correction.
To provide this supervision, we construct 642K exactly duration-matched English--Chinese speech pairs by translating and synthesizing target speech on source-defined temporal scaffolds.
Training combines a progressive supervised curriculum with verifier-guided reinforcement learning to balance translation quality and temporal alignment.
On a 4,353-utterance test set spanning read and spontaneous speech, \model{} increases the proportion of translations within 2\% of the source duration from 25.6\% to 61.5\% and pause-alignment F1 from 0.118 to 0.550 compared with \uniss{}, while maintaining competitive translation quality.
Our analysis shows that aligned supervision establishes timeline following, while reinforcement learning refines temporal precision and partially offsets the translation-quality cost of strict synchronization.
Together, these results demonstrate the effectiveness of our data and training framework for timeline-preserving S2ST. Audio samples, code, datasets, and pretrained models are available on our project page at \href{https://zhishengzheng.com/transdubber/}{TransDubber}.

\end{abstract}

\section{Introduction}
\label{sec:introduction}

In a dubbed scene, silence can be part of the dialogue. A character may pause to breathe, hesitate, or wait for a reaction; a translation that speaks through that interval can disrupt the timing of the scene even when its words are accurate. Psycholinguistic research shows that prosodic boundaries guide listeners' online interpretation of sentence structure~\citep{steinhauer1999brain}, motivating attention to how speech is organized as well as what it conveys. In audiovisual translation, phrase and pause alignment is also tied to the visual context, with different synchronization requirements for on-screen and off-screen speech~\citep{virkar2022prosodic}. We therefore focus on settings in which translated speech must fit an existing scene or event sequence. 

\begin{figure}[htb]
    \centering
    \captionsetup{skip=2pt}
    \includegraphics[width=\linewidth]{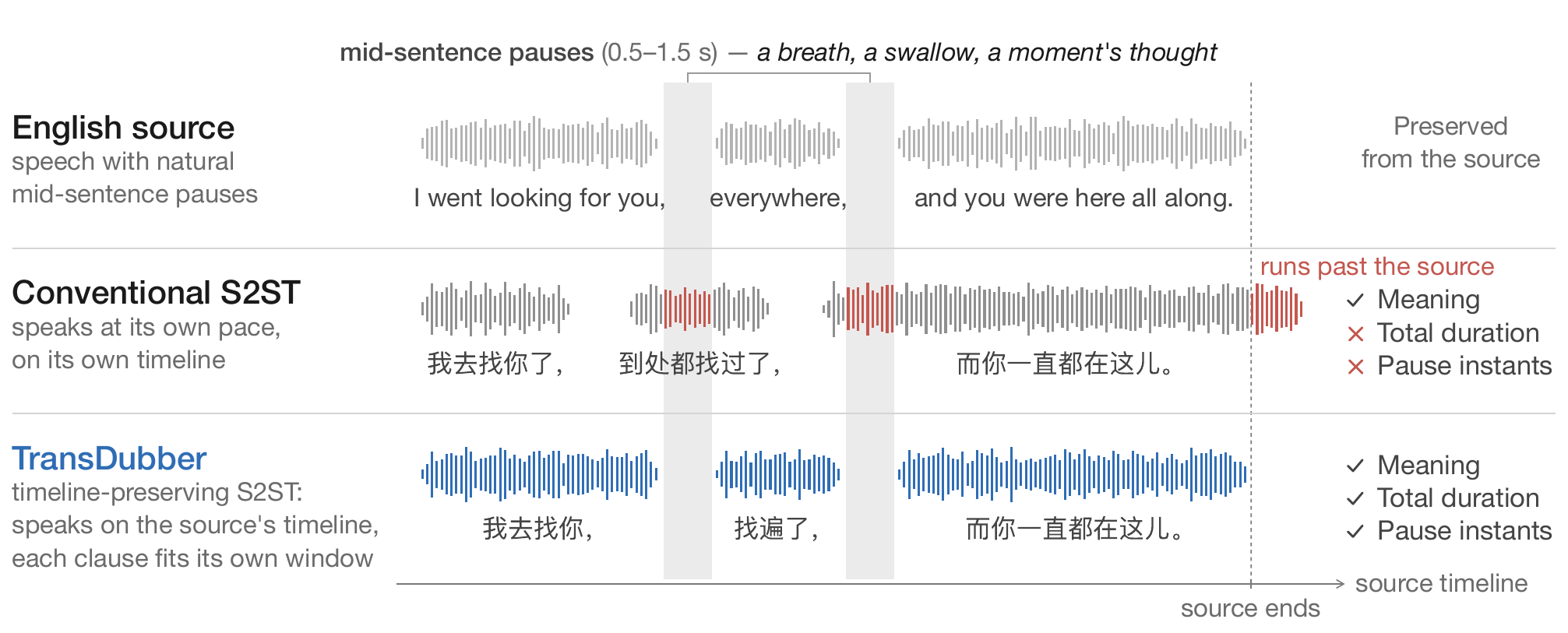}
    \caption{Illustration of timeline-preserving speech translation. Conventional S2ST may shift pauses or overrun the source endpoint; \model instead preserves meaning while aligning translated speech with the source speaking windows, salient pauses, and utterance endpoint.
    }
    \label{fig:novelty}
\end{figure}

Modern speech-to-speech translation (S2ST) systems increasingly preserve semantic content, speaker identity, and expressive characteristics~\citep{jia2019translatotron,lee2022direct,seamless2023expressive,barrault2023seamlessm4t,cheng2025uniss}. Temporal constraints have also been studied at both the translation and speech-generation levels: isometric MT encourages source--target text-length agreement~\citep{lakew2022isometric}, while duration-based translation uses source-speech duration to guide translated text~\citep{won2025duration}. Prior work addresses internal temporal structure: isochrony-aware MT jointly translates and projects pause markers~\citep{tam2022isochrony}, and TransVIP conditions speech generation on duration and frame-level voice activity through an explicit isochrony-control module~\citep{le2024transvip}. Matching the source endpoint alone does not determine this internal timeline: two translations can have exactly the same duration while placing pauses and speaking regions at entirely different moments. Building on these efforts, we study \emph{timeline-preserving S2ST}, in which target speech aims to preserve the source utterance endpoint and salient internal speech--silence events on a shared absolute timeline, as illustrated in Fig.~\ref{fig:novelty}. Our focus is on how an autoregressive S2ST model can learn this behavior directly from target waveforms aligned to the source timeline, without an inference-time frame-level isochrony controller or post-hoc padding, truncation, or global time stretching. This is an application-specific constraint, rather than an assumption that source-like timing is universally optimal: studies of professional dubbing emphasize translation quality and vocal naturalness and suggest a more conditional role for timing constraints~\citep{brannon2023dubbing}.

Learning this behavior requires supervision that standard parallel speech corpora do not provide. Independently recorded translations naturally follow the phrasing and rhythm of the target language rather than the speech-activity pattern of the source. To create the missing supervision, we develop a scalable source-canvas construction pipeline. We extract a speech-and-pause scaffold from each source utterance, translate individual speech regions under local duration budgets while retaining utterance-level semantic context, synthesize speaker-conditioned target speech, and place the accepted segments back into their corresponding source windows. The resulting targets preserve the source endpoint and selected internal pauses by construction, providing explicit supervision for both pause-bearing and pause-free timelines.

On this data, we train \model with a progressive multi-stage curriculum. The model first acquires pause-aware recognition and translation, then learns natural target-speech generation, and finally learns to generate translated speech on the source timeline. We subsequently apply verifier-guided policy optimization with separate objectives for translation quality, global duration agreement, and local pause alignment. Rather than collapsing these signals into a single reward before optimization, we normalize them independently and assign semantic feedback to translated-text tokens while assigning temporal feedback to speech-codec tokens. On our English-to-Chinese test set, \model improves Dur@2\% from 25.6\% to 61.5\% and pause F1 from 0.118 to 0.550 relative to UniSS, with a similar COMET score.



Our contributions are as follows:
\begin{itemize}
    \item \textbf{Timeline preservation as a learned generation behavior.}
    We formulate timeline-preserving S2ST and train \model to reproduce the source endpoint and salient internal pauses directly through autoregressive generation, without an inference-time frame-level timing controller or post-hoc temporal correction.
    
    \item \textbf{A 642K-pair timeline-preserving corpus.}
    Our scalable pipeline combines duration-budgeted translation, constrained speaker-conditioned synthesis, and source-canvas rendering to produce exactly duration-matched English--Chinese speech pairs with 298K supervised pause spans.

    \item \textbf{Progressive training with fine-grained RL.}
    We progress from pause-aware text grounding to natural speech generation and timeline-preserving fine-tuning, followed by verifier-guided RL. Quality, duration, and pause rewards are normalized independently, with semantic and temporal feedback assigned to text and speech tokens, respectively.

    \item \textbf{State-of-the-art timeline-preserving S2ST.}
    \model substantially outperforms strong S2ST baselines in strict duration fidelity and local pause alignment while maintaining competitive translation quality.
\end{itemize}

\section{Constructing Timeline-Preserving Supervision}
\label{sec:data}

We construct the supervision data in three stages. Stage~1 derives a source-side speech-and-pause scaffold from each English utterance, transcribes the resulting speech regions, and separately produces a natural Chinese translation. Stage~2 uses the natural translation to construct complete Chinese speech at its natural speaking rate. Stage~3 reuses the Stage~1 temporal scaffold and region-level English transcripts, but translates these transcripts again under local duration constraints before synthesizing and rendering the target speech on the source timeline.

Stages 1 and 2 are constructed from the 8--25\,s source pool drawn from Emilia-EN~\citep{he2024emilia} and GigaSpeech~\citep{chen2021gigaspeech}. Stage~3 uses eligible utterances from this pool and additionally introduces 2--8\,s utterances to improve short-form coverage. Detailed construction and filtering settings are provided in Appendix~\ref{app:data-pipeline}.

\subsection{Stage 1: text grounding}

\paragraph{Source-derived temporal scaffold.}
For each source waveform, we first apply source-side VAD~\citep{team2021silero} to identify speech spans and the internal non-speech gaps between adjacent spans. Internal gaps of at least 0.5\,s are selected as pause boundaries, and the utterance is partitioned into speech regions only at these selected gaps. The pause locations are therefore derived directly from the source waveform rather than from punctuation or translated text.

Segment-level ASR~\citep{qwen2026asr} is applied to the resulting speech regions, producing English transcripts \(e_{1:N}\). For utterances with selected internal pauses, the region transcripts are joined using a dedicated textual marker. Each marker corresponds to a selected source-side pause interval.
\begin{equation*}
    e^{\mathrm{pause}} =
e_1\;\texttt{<pause>}\;e_2\;\cdots\;\texttt{<pause>}\;e_N
\end{equation*}
Utterances without a selected internal pause contain no \texttt{<pause>} marker.

Separately, Hy-MT2-30B-A3B~\citep{zheng2026hymt2} translates the complete English utterance into a natural Chinese sentence \(z^{\mathrm{nat}}\). This translation is generated at the utterance level without a duration budget, character-count constraint, region-level alignment requirement, or pause-placement constraint. Unlike the English ASR representation, \(z^{\mathrm{nat}}\) contains no \texttt{<pause>} token.

Stage~1 therefore yields ASR pairs \((x,e^{\mathrm{pause}}) \in \mathcal{D}_{\mathrm{ASR}}\) and AST pairs \((x,z^{\mathrm{nat}}) \in \mathcal{D}_{\mathrm{AST}}\). The temporal scaffold and \(e_{1:N}\) are retained for Stage~3, while \(z^{\mathrm{nat}}\) also serves as the synthesis text for Stage 2.


\subsection{Stage 2: natural-pace speech supervision}

Stage~2 uses the natural Chinese translation \(z^{\mathrm{nat}}\) from Stage~1 as synthesis text. For each source utterance, VoxCPM2~\citep{voxcpm2technical} synthesizes a complete Chinese waveform \(y^{\mathrm{nat}} = \operatorname{VoxCPM2}(z^{\mathrm{nat}},r_x)\), conditioned on source-speaker reference audio \(r_x\).

The synthesized waveform follows the natural phrasing, rhythm, and duration of the Chinese translation. No source-duration constraint, segment-level duration budget, time-scale modification, or source-canvas rendering is applied.

\begin{figure}[htb]
    \centering
    \includegraphics[width=1.0\linewidth]{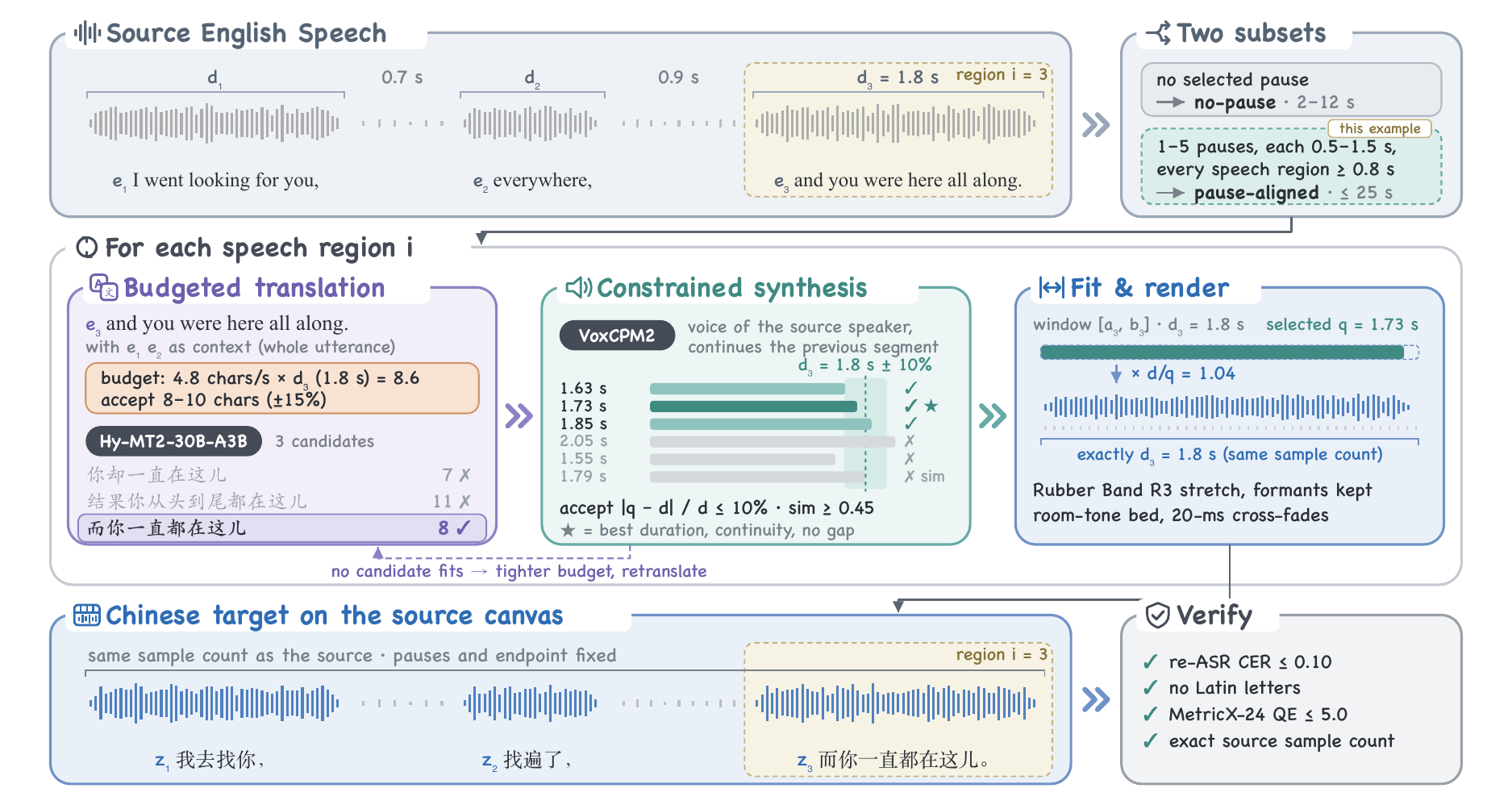}
    \caption{Stage 3 construction pipeline for timeline-preserving speech supervision.}
    \label{fig:data_stage3_canvas}
\end{figure}

\subsection{Stage 3: Timeline-Preserving Speech Supervision}

Stage 3 reuses the Stage 1 temporal scaffold and English region transcripts, but generates new duration-constrained translations rather than using $z^{\mathrm{nat}}$ (Fig.~\ref{fig:data_stage3_canvas}). It includes a \emph{no-pause} subset without selected internal pauses and a \emph{pause-aligned} subset preserving selected pauses of approximately 0.5--1.5\,s.

\paragraph{Duration-budgeted translation.}
Each region transcript $e_i$ is translated under a duration-dependent character budget, with the full transcript $e_{1:N}$ as semantic context. When synthesis cannot satisfy the duration constraint, we refine the budget and retranslate the region.

\paragraph{Constrained synthesis and source-canvas rendering.}
VoxCPM2 generates multiple source-speaker-conditioned candidates, selected for duration compatibility, speaker similarity, and cross-segment continuity. Only near-duration candidates are eligible for bounded time-scale correction; an utterance is discarded if any region lacks an acceptable candidate. Accepted segments are fitted to their assigned source windows, while audio outside these windows is retained. The resulting waveform preserves the selected source pause intervals and utterance endpoint and has exactly the same sample count as the source. These corrections are used only to construct training targets.

\paragraph{Verification and retained data.}
Quality and timing checks yield 642K speech pairs with 298K supervised pause spans (Table~\ref{tab:training-data}).

\begin{table*}[htb]
  \centering
  \caption{Retained training data by stage and supervision type. Source hours are summed across tasks and stages, not unique corpus hours. Stage 1 pause markers annotate source transcripts; only Stage 3 provides target-waveform timing supervision.}

  \label{tab:training-data}
  \scriptsize
  \begin{tabular}{llrrrll}
    \toprule
    Stage & Data type & \# Samples & Src. h  & Tgt. h
          & Target timing & Pause supervision \\
    \midrule
    1 & ASR text
      & 928,661 & 3,014 & --
      & Not applicable
      & Source-side markers only \\
    1 & AST text
      & 883,563 & 2,854 & --
      & Not applicable
      & None in Chinese target \\
    2 & Natural-pace S2ST
      & 806,407 & 2,610 & 2,431
      & Unconstrained natural timing
      & No target-time labels \\
    3 & Timeline no-pause
      & 412,945 & 687 & 687
      & Exact source duration
      & No selected pause spans \\
    3 & Timeline pause-aligned
      & 229,331 & 682 & 682
      & Exact source duration
      & All selected pauses retained \\
    \midrule
    \multicolumn{2}{l}{Stage-summed total}
      & 3,260,907 & 9,847 & 3,800
      & -- & -- \\
    \bottomrule
  \end{tabular}
\end{table*}

\section{Timeline-Preserving Training}
\label{sec:method}

\subsection{Model and sequence representation}

\model is initialized from Qwen3-ASR-1.7B~\citep{qwen2026asr}, a speech recognition model that maps input speech to text but does not natively support translation or speech generation. We extend its vocabulary with 6,561 discrete speech-codec symbols~\citep{du2025cosyvoice}, representing speech at a frame rate of 25~Hz, and train the model to generate translated text and target speech in addition to the source-language ASR transcript.

\begin{figure}[htb]
    \centering
    \includegraphics[width=0.9\linewidth]{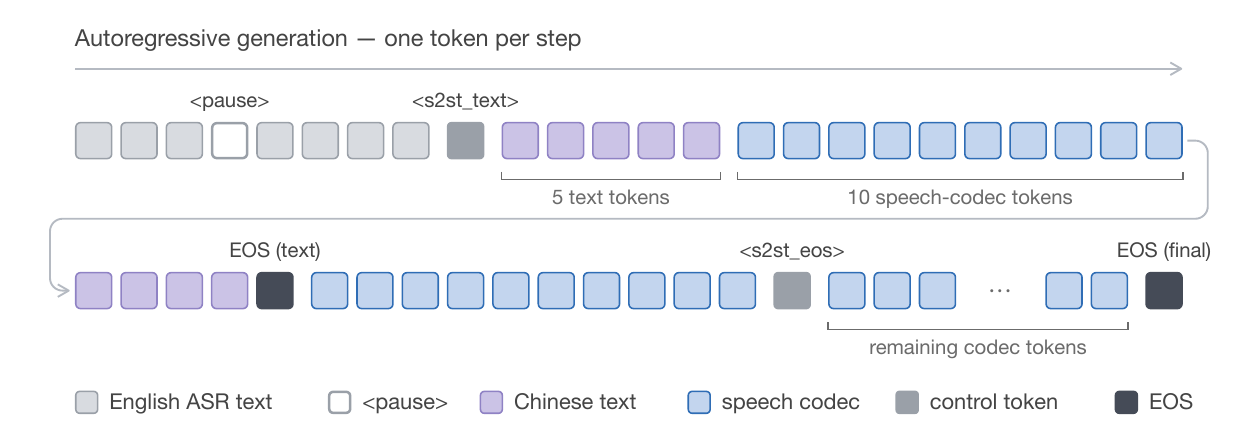}
    \caption{Autoregressive decoding in \model. A pause-marked English transcript precedes Chinese text and speech-codec tokens, interleaved in blocks of five text and ten speech tokens. After text completion, speech generation continues until the final end-of-sequence token.}
    \label{fig:model_inference}
\end{figure}

Given a source utterance, \model first autoregressively generates the ASR transcript, followed by the translated text and target-speech codec tokens in an interleaved fashion, as illustrated in Fig.~\ref{fig:model_inference}. Following~\citet{wu2025towards}, we adopt a fixed 5:10 schedule, where every five text tokens are followed by ten speech tokens. Once the translated text is complete, no additional text padding tokens are inserted; the model continues generating speech tokens until the end-of-sequence token is produced. A flow-matching-based speech decoder~\citep{du2025cosyvoice} then maps the generated discrete speech-token sequence to the corresponding raw waveform.


\subsection{Stage 1: text grounding}
The first stage trains pause-aware ASR and English-to-Chinese text translation. Source transcripts include explicit pause markers when valid pauses are present. It does not yet require speech generation or source-timeline following.

\subsection{Stage 2: natural-pace speech-generation warm-up}
\label{sec:train_stage2}

Stage~2 trains the model to generate translated text and target-speech codec tokens using naturally paced target speech. The output follows the interleaved text--speech sequence format described above, but the target waveforms are not constrained to match the source duration or its internal pause locations. This stage therefore introduces target-speech generation and sequence termination without explicitly imposing source-timeline constraints.

\subsection{Stage 3: timeline-preserving supervised fine-tuning}
\label{sec:train_stage3}

Stage~3 fine-tunes the model on the timeline-aligned target waveforms
constructed in Section~\ref{sec:data}. Each target matches the source
utterance duration and retains the selected internal pause intervals,
when present. Maximum-likelihood training on these targets provides
supervision for placing translated speech within the source speaking
windows and terminating near the source endpoint.

\subsection{Stage 4: Verifier-guided policy optimization}
\label{sec:rl}

After Stage~3, we further optimize the policy with GSPO~\citep{zheng2025gspo} using verifier-based rewards for translation quality and temporal alignment. We sample RL prompts from the same source-utterance pool used for Stage~3, including both no-pause and pause-aligned examples. For each source prompt $x$, the policy samples a group of $G$ candidates $\{y_i\}_{i=1}^{G}$. Each candidate is evaluated by three sequence-level rewards: translation quality $R_{\mathrm{qual}}$, global duration agreement $R_{\mathrm{dur}}$, and pause alignment $R_{\mathrm{pause}}$.

Importantly, we do not combine these rewards into a scalar before advantage estimation. Instead, each reward is normalized independently within the rollout group and optimized using the token subset most directly associated with that reward.

\paragraph{Verifier rewards.}
We measure translation quality using MetricX-QE~\citep{juraska-etal-2024-metricx} in its reference-free mode. Given the dataset-provided source transcript $s_x$ and the generated translation $\hat{t}_y$, MetricX-QE assigns a score in $[0,25]$, with lower values indicating better translation quality. We convert this score into a higher-is-better reward in $[0,1]$ via \(R_{\mathrm{qual}}(x,y) = 1 - \operatorname{MetricX\text{-}QE}(s_x,\hat{t}_y) / {25}\).

Using the dataset transcript, rather than the model's own ASR output, prevents recognition errors from directly affecting the quality verifier.

\begin{figure}[htb]
    \centering
    \begin{subfigure}[t]{0.48\linewidth}
        \centering
        \includegraphics[width=\linewidth]{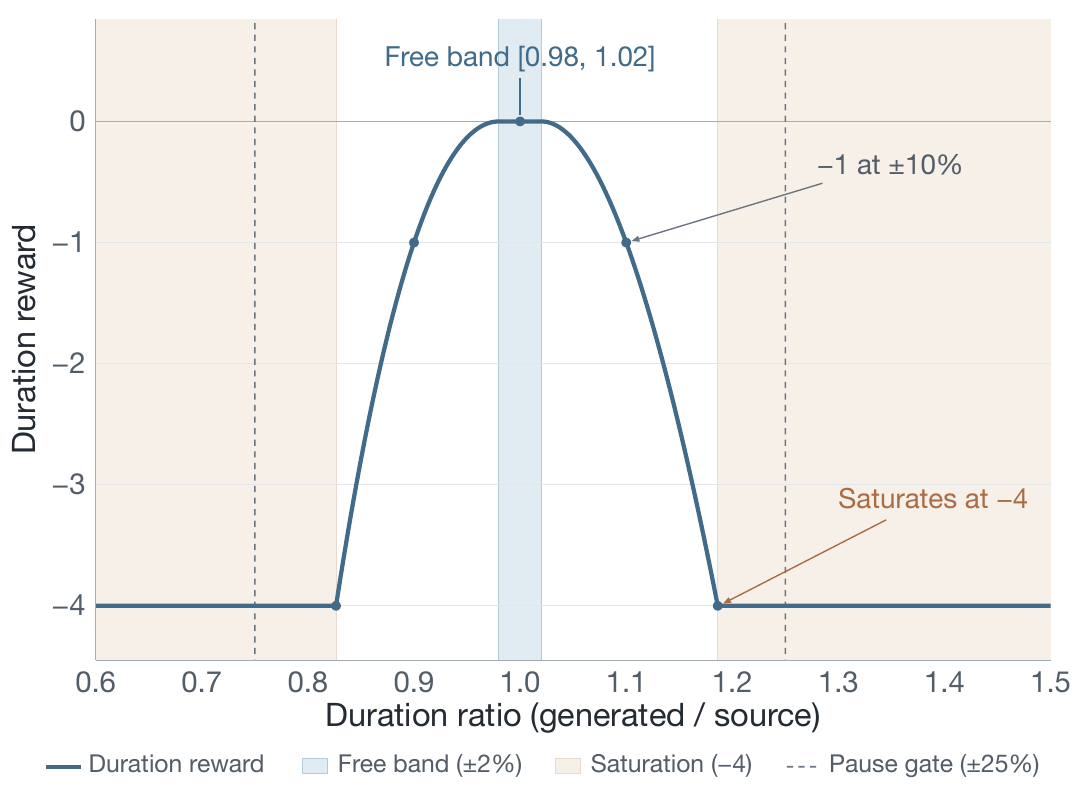}
        \caption{Global duration reward.}
        \label{fig:duration_reward}
    \end{subfigure}
    \hfill
    \begin{subfigure}[t]{0.51\linewidth}
        \centering
        \includegraphics[width=\linewidth]{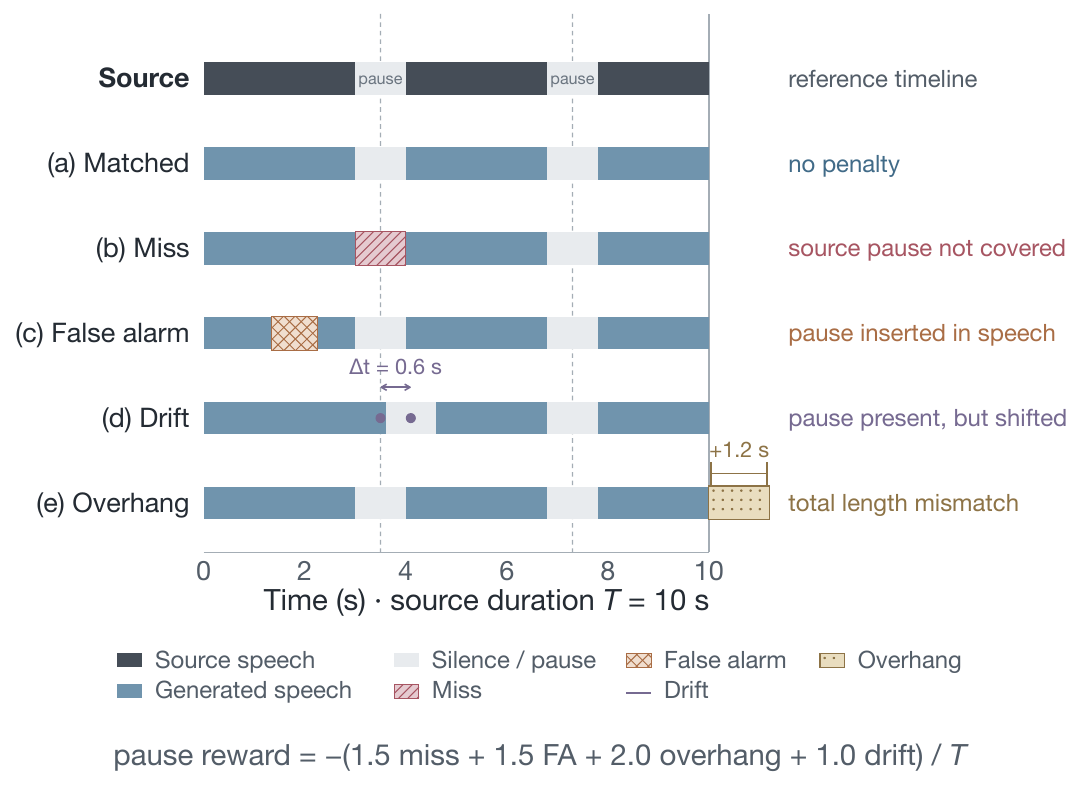}
        \caption{Timeline-based pause reward.}
        \label{fig:pause_reward}
    \end{subfigure}
    \caption{Temporal rewards for Stage 4 reinforcement learning. (a) The duration reward is zero within ±2\% of source duration and saturates at -4. (b) The pause reward penalizes missed or spurious silence, shifted pause centers, and endpoint mismatch.}
    \label{fig:reward}
\end{figure}

For duration alignment, let $T_x$ and $T_y$ denote the source and generated speech durations, respectively. The duration reward is defined from the ratio $\rho=T_y/T_x$. As illustrated in Fig.~\ref{fig:duration_reward}, it assigns no penalty within $\pm 2\%$ of the source duration, decreases quadratically with the normalized log-distance outside this region, reaches $-1$ at $\pm 10\%$, and saturates at $-4$ for large mismatches.

The pause reward compares source and generated speech--silence occupancy on a shared timeline. It penalizes four types of temporal error: uncovered source pauses, pauses inserted during source speech, pause-center displacement, and endpoint overhang. A small temporal collar allows minor boundary differences between corresponding pause intervals. To prevent a temporally mis-scaled candidate from receiving a favorable local-alignment reward, we additionally enforce $\lvert \rho-1\rvert>0.25 \Longrightarrow R_{\mathrm{pause}}(x,y)\leq -1.
  \label{eq:pause-duration-gate}
$
The exact duration function, pause-interval construction, and penalty weights are provided in Appendix~\ref{app:reward-details}.

\paragraph{Component-specific credit assignment.}
The quality reward is assigned only to the tokens of the generated text translation, excluding speech-codec and special tokens. In contrast, the duration and pause rewards are assigned only to generated speech-codec tokens.

Let $M_{\mathrm{text}}$ and $M_{\mathrm{speech}}$ denote the corresponding token masks, and let $A_c$ be the independently group-normalized advantage for component $c$. We optimize
\begin{align}
  \mathcal{L}
  =
  \frac{1}{\lambda_q+\lambda_p+\lambda_d}\Big(
  &\lambda_q\,\mathcal{L}_{\mathrm{GSPO}}
    (A_{\mathrm{qual}},M_{\mathrm{text}})
  +
  \lambda_p\,\mathcal{L}_{\mathrm{GSPO}}
    (A_{\mathrm{pause}},M_{\mathrm{speech}})
  \nonumber\\
  &+
  \lambda_d\,\mathcal{L}_{\mathrm{GSPO}}
    (A_{\mathrm{dur}},M_{\mathrm{speech}})
  \Big)
  +
  \beta\,
  D_{\mathrm{KL}}^{M_{\mathrm{text}}\cup M_{\mathrm{speech}}}
  \left(
    \pi_{\theta}\,\|\,\pi_{\mathrm{SFT}}
  \right).
  \label{eq:component-gspo}
\end{align}

Each GSPO term uses a sequence-level importance ratio computed over its own token mask. Details of reward normalization, advantage clipping, importance-ratio clipping, and rollout filtering are reported in Appendix~\ref{app:reward-details}.

\section{Experiments}
\label{sec:experiments}

\subsection{Experimental Setup}
\label{sec:setup}

\paragraph{Evaluation data.}
We construct an English-to-Chinese evaluation set by selecting 4,353 real speech utterances from the CoVoST2 test set \citep{wang2021covost2} and GigaSpeech test set \citep{chen2021gigaspeech}. The resulting set contains 1,439 CoVoST2 utterances, which are predominantly read speech and come with human Chinese references, and 2,914 GigaSpeech utterances, which consist of spontaneous speech and are paired with Chinese references generated by Gemini 3.1 Pro and subsequently manually reviewed by experts for correctness. The source utterances have an average duration of 8.9 seconds and a maximum duration of 19.4 seconds. 

\paragraph{Systems.}
We compare \textbf{\model{}} against five representative baselines spanning cascaded pipelines, specialized end-to-end S2ST models, expressive S2ST systems, and general-purpose multimodal models. \textbf{Cascaded} is a strong three-stage pipeline consisting of Qwen3-ASR-1.7B~\citep{qwen2026asr} for English speech recognition, Hy-MT2-30B-A3B~\citep{zheng2026hymt2} for English-to-Chinese text translation, and VoxCPM2~\citep{voxcpm2technical} for Chinese speech synthesis. \textbf{Seamless-L}~\citep{barrault2023seamlessm4t} is a large multilingual end-to-end speech translation model from the SeamlessM4T family, while \textbf{Seamless-Ex}~\citep{seamless2023expressive} refers to SeamlessExpressive, an end-to-end S2ST model designed to preserve expressive and paralinguistic information. \textbf{Qwen2.5-Omni-7B}~\citep{jin2025Omni} is a general-purpose multimodal model supporting direct speech input and speech output. \textbf{\uniss}~\citep{cheng2025uniss} is a unified expressive S2ST baseline. TransVIP~\citep{le2024transvip} is closely related but is not included because its published system targets French--English rather than our English--Chinese setting.


\paragraph{Evaluation metrics and protocol.}
We evaluate the proposed system as follows:
\begin{itemize}
    \item \textbf{Translation quality and speaker preservation.} We report MetricX-QE, MetricX-ref~\citep{juraska-etal-2024-metricx} and COMET score~\citep{rei2020comet}. We also report speaker similarity (SIM; ~\citealp{chen2022wavlm}) to evaluate speaker preservation.

    \item \textbf{Isochrony (global duration fidelity).} We report $\durat{2\%}$, $\durat{5\%}$, and $\durat{10\%}$, which measure the percentage of utterances whose relative duration error falls within the corresponding threshold. 
    
    \item \textbf{Pause alignment (local timeline fidelity).} We report pause F1@0.4 and onsetMedAE@0.4 for event-level pause alignment, and pause-span tIoU and SER for span-level temporal alignment. 
    
    \item \textbf{Human evaluation.} We report 1--5 ratings for translation quality (Qual.) and speech naturalness (Nat.).
    
\end{itemize}

See Appendix~\ref{app:timing_metrics} for detailed definitions of the isochrony and pause-alignment metrics, and Appendix~\ref{app:human_eval} for the human-evaluation protocol. Unless otherwise stated, all automatic translation-quality scores, including COMET and MetricX-QE, are computed on the intended-text track, using each system's target-language translation text rather than an ASR transcription of its generated speech. MetricX-QE is used for RL optimization, whereas COMET with the \texttt{wmt22-comet-da} checkpoint is held out from training and used as an independent audit metric.

\paragraph{Decoding.}
Unless otherwise stated, \model{} decoding uses nucleus sampling with $p=0.85$, temperature $0.85$, no top-$k$ cutoff, a maximum of 896 new tokens, and seed 42.

\subsection{Main Results}
\label{sec:results}


Table~\ref{tab:main} demonstrates the benefit of explicitly learning source-timeline preservation. \model{} achieves $\durat{2\%}$ and $\durat{5\%}$ of 61.5\% and 92.2\%, respectively, compared with 25.6\% and 56.1\% for \uniss{}, the strongest baseline on these metrics. Across Cascaded, Seamless-L, Seamless-Ex, and Qwen2.5-Omni-7B, $\durat{2\%}$ remains between 3.7\% and 5.4\%. The improvements extend beyond the utterance endpoint to internal pause placement: \model{} reaches 0.550 pause F1, compared with 0.118 for \uniss{} and at most 0.080 for the other baselines. It also obtains the best tIoU and SER, at 0.322 and 0.128, respectively. 


\begin{table*}[htb]
\centering
\caption{English-to-Chinese speech-to-speech translation on 4,353 CoVoST2 and GigaSpeech test utterances. Dur@$\epsilon$ denotes the percentage within $\epsilon$ of source duration. onMedAE denotes onsetMedAE@0.4. Human ratings are on a 1--5 scale.}

\label{tab:main}
\small
\setlength{\tabcolsep}{3.5pt}
\renewcommand{\arraystretch}{1.08}
\begin{adjustbox}{max width=\textwidth}
\begin{tabular}{lcccccccccccc}
\toprule
\multirow{2}{*}{System} & \multirow{2}{*}{SIM$\uparrow$}
& \multicolumn{2}{c}{Translation Quality}
& \multicolumn{3}{c}{Isochrony}
& \multicolumn{4}{c}{Pause Alignment}
& \multicolumn{2}{c}{Human Evaluation} \\
\cmidrule(lr){3-4}\cmidrule(lr){5-7}\cmidrule(lr){8-11}\cmidrule(lr){12-13}
& & COMET$\uparrow$ & MetricX-QE$\downarrow$
& $\durat{2\%}$$\uparrow$ & $\durat{5\%}$$\uparrow$ & $\durat{10\%}$$\uparrow$
& F1$\uparrow$ & tIoU$\uparrow$ & SER$\downarrow$ & onMedAE$\downarrow$
& Qual.$\uparrow$ & Nat.$\uparrow$ \\
\midrule
Cascaded
& 0.451 & 0.854 & 2.67
& 5.4 & 13.6 & 27.3
& 0.076 & 0.071 & 0.402 & 0.213
& 3.97 & 3.93 \\
\midrule
Seamless-L
& 0.029 & 0.705 & 6.79
& 3.7 & 9.5 & 19.5
& 0.002 & 0.008 & 0.407 & 0.533
& 1.83 & 1.10 \\
Seamless-Ex
& 0.261 & 0.718 & 5.99
& 4.0 & 10.7 & 21.9
& 0.062 & 0.059 & 0.395 & 0.230
& 2.20 & 1.53 \\
Qwen2.5-Omni-7B
& 0.020 & \textbf{0.809} & \textbf{3.88}
& 4.0 & 11.1 & 21.5
& 0.080 & 0.060 & 0.446 & 0.211
& 3.63 & \textbf{3.97} \\
\uniss
& 0.293 & 0.784 & 5.65
& 25.6 & 56.1 & 83.3
& 0.118 & 0.108 & 0.288 & 0.266
& 3.13 & 2.87 \\
\midrule

\rowcolor{OursRow}
\model
& \textbf{0.464} & 0.788 & 4.12
& \textbf{61.5} & \textbf{92.2} & \textbf{98.7}
& \textbf{0.550} & \textbf{0.322} & \textbf{0.128} & \textbf{0.162}
& \textbf{3.63} & 3.47 \\
\bottomrule
\end{tabular}
\end{adjustbox}
\end{table*}

These temporal gains are accompanied by competitive translation quality. \model{} achieves 0.788 COMET and 4.12 MetricX-QE, outperforming both Seamless variants; relative to \uniss{}, COMET is comparable (0.788 vs.\ 0.784), while MetricX-QE improves from 5.65 to 4.12. Human evaluation provides complementary evidence: on the 1--5 scales, \model{} receives 3.63 for translation quality and 3.47 for speech naturalness, exceeding \uniss{} (3.13 and 2.87) and both Seamless variants on both dimensions. Its translation-quality rating also matches Qwen2.5-Omni-7B at 3.63, despite the latter's higher COMET score (0.809). Cascaded still achieves higher automatic and human-rated translation quality (0.854 COMET and 3.97 Qual.), and both Cascaded and Qwen2.5-Omni-7B receive higher naturalness ratings (3.93 and 3.97, respectively). 



Table~\ref{tab:progression} further shows how these capabilities emerge through the training curriculum. Stage~1 establishes text translation, while Stage~2 introduces natural-pace speech generation but provides limited temporal fidelity, with only 5.2\% $\durat{2\%}$ and 0.099 pause F1. Its COMET score of 0.823 is computed only on the 73.7\% of samples that successfully complete generation. Timeline-preserving SFT in Stage~3 produces the main gains in source-timeline following, raising $\durat{2\%}$ to 48.6\% and pause F1 to 0.518. Stage~4 then refines this behavior: $\durat{2\%}$ and $\durat{5\%}$ increase to 61.5\% and 92.2\%, respectively, pause F1 reaches 0.550, and SER decreases from 0.198 to 0.128. These temporal improvements are accompanied by an increase in COMET from 0.772 to 0.788. 

\begin{table*}[htb]
  \centering
  \caption{Capability acquisition across \model{}’s training stages. Stage 1 generates text only.}
  \label{tab:progression}
  \footnotesize
  \begin{threeparttable}
  \begin{adjustbox}{max width=\textwidth}
  \begin{tabular}{L{1.4cm}L{4.1cm}cccccc}
    \toprule
    Stage
    & Capability
    & COMET $\uparrow$
    & $\durat{2\%}$ $\uparrow$
    & $\durat{5\%}$ $\uparrow$
    & Pause F1 $\uparrow$
    & tIoU $\uparrow$
    & SER $\downarrow$ \\
    \midrule

    Stage 1
    & ASR and text translation
    & \textbf{0.823} & -- & -- & -- & -- & -- \\

    \midrule

    Stage 2
    & Natural-pace speech translation
    & 0.823$^{\dagger}$ & 5.2 & 14.2 & 0.099 & 0.086 & 0.420 \\

    Stage 3
    & Timeline-preserving translation
    & 0.772 & 48.6 & 83.3 & 0.518 & 0.293 & 0.198 \\

    Stage 4
    & Quality and timing refinement
    &  0.788 & \textbf{61.5} & \textbf{92.2} & \textbf{0.550} & \textbf{0.322} & \textbf{0.128} \\




    \bottomrule
  \end{tabular}
  \end{adjustbox}

  \begin{tablenotes}[flushleft]
    \footnotesize
    \item[$\dagger$] Computed only on the 73.7\% of samples that successfully completed generation.
  \end{tablenotes}
  \end{threeparttable}
\end{table*}

\subsection{Ablation Studies and Analysis}
\label{sec:analysis}

\paragraph{Role of pause markers.}
We intervene on the \texttt{<pause>} markers in the ASR prefix to examine whether the pause-aware representation guides target-timeline generation (Table~\ref{tab:pause_causal_intervention}). Shifting the markers forward by three tokens reduces pause F1 from 0.550 to 0.476 while leaving the number of generated pauses almost unchanged (0.89 vs.\ 0.88 per utterance). Shuffling their positions further reduces F1 to 0.369. Thus, the marker locations affect pause alignment beyond simply controlling how often the model pauses. Removing the markers produces the largest degradation: pause F1 falls to 0.231, tIoU decreases from 0.322 to 0.156, and the generated pause count approximately halves to 0.44 per utterance. Nevertheless, $\durat{2\%}$ remains 58.3\%, and COMET improves slightly to 0.795. These interventions show that both the presence and placement of pause markers contribute to local timeline fidelity, which can deteriorate substantially even when translation quality and utterance-level duration agreement remain comparatively stable.

\begin{table*}[htbp]
\centering
\caption{Pause-marker interventions in \model. ASR inputs contain an average of 0.89 \texttt{<pause>} markers per utterance.}
\label{tab:pause_causal_intervention}
\resizebox{\textwidth}{!}{%
\begin{tabular}{lcccccccc}
\toprule
& COMET$\uparrow$
& MetricX-QE$\downarrow$
& $\durat{2\%}\uparrow$
& $\durat{5\%}\uparrow$
& Pause F1$\uparrow$
& tIoU$\uparrow$
& onMedAE$\downarrow$
& \textbf{Pauses / utt.} \\
\midrule

\rowcolor{OursRow}
\textbf{No intervention}
& 0.788 & 4.12 & \textbf{61.5} & \textbf{92.2} & \textbf{0.550} & \textbf{0.322} & \textbf{0.162} & 0.89 \\

Shift $+3$ tokens
& 0.784 & 4.24 & 60.7 & 91.5 & 0.476 & 0.278 & 0.174 & 0.88 \\

Shuffle
& 0.786 & 4.17 & 58.8 & 90.2 & 0.369 & 0.227 & 0.174 & 0.81 \\

Remove
& \textbf{0.795} & \textbf{3.91} & 58.3 & 89.1 & 0.231 & 0.156 & 0.192 & 0.44 \\

\bottomrule
\end{tabular}%
}
\end{table*}

\paragraph{Natural-pace speech-generation warm-up.}
Table~\ref{tab:stage_ablation} compares curricula with and without Stage~2 before timeline-preserving SFT. Including the natural-pace warm-up modestly improves translation quality, increasing COMET from 0.768 to 0.772 and lowering both MetricX scores. Its main benefit is stronger text--speech consistency: average CER between generated text and re-transcribed speech decreases from 5.14\% to 4.24\%, while the fraction of outputs with CER at least 30\% falls from 3.81\% to 2.92\%. Temporal performance changes little, with nearly identical $\durat{10\%}$ (94.9\% vs.\ 94.8\%) and a slight increase in pause F1 (0.511 to 0.518). These results suggest that the warm-up primarily strengthens speech realization rather than timeline alignment.

\begin{table*}[htbp]
  \centering
  \caption{Ablation on progressive capability acquisition. The intermediate natural-pace speech translation stage ($S_2$) improves both translation quality and speech intelligibility.}
  \label{tab:stage_ablation}
  \small
  \begin{threeparttable}
  \begin{adjustbox}{max width=\textwidth}
  \begin{tabular}{L{2.5cm}ccccccc}
    \toprule
    \multirow{2}{*}{Training Curriculum}
    & \multicolumn{3}{c}{Translation Quality}
    & \multicolumn{2}{c}{Speech Intelligibility$^\dagger$}
    & \multirow{3}{*}{$\durat{10\%}$ $\uparrow$}
    & \multirow{3}{*}{Pause F1 $\uparrow$} \\
    \cmidrule(lr){2-4}
    \cmidrule(lr){5-6}

    & MetricX-ref $\downarrow$
    & MetricX-QE $\downarrow$
    & COMET $\uparrow$
    & Avg. CER (\%) $\downarrow$
    & CER $\geq 30\%$ (\%) $\downarrow$
    & & \\
    \midrule

    $S_1 \!\rightarrow\! S_3$
    & 4.35
    & 4.49
    & 0.768
    & 5.14
    & 3.81
    & 94.9
    & 0.511 \\

    $\mathbf{S_1 \!\rightarrow\! S_2 \!\rightarrow\! S_3}$
    & \textbf{4.21}
    & \textbf{4.38}
    & \textbf{0.772}
    & \textbf{4.24}
    & \textbf{2.92}
    & 94.8
    & \textbf{0.518} \\

    \bottomrule
  \end{tabular}
  \end{adjustbox}

  \begin{tablenotes}[flushleft]
    \scriptsize
    \item[$\dagger$] CER is computed between the Paraformer transcription of the generated speech and the corresponding generated text.
  \end{tablenotes}
  \end{threeparttable}
\end{table*}

\paragraph{Contributions of the reward components.}
Table~\ref{tab:reward-ablation} shows the trade-off among the reward components. Quality-only optimization gives the highest translation scores but substantially weakens temporal alignment. The pause reward yields the highest pause F1 and tIoU, whereas combining the temporal rewards improves strict duration matching. Adding the quality reward to this combination raises COMET from 0.772 to 0.788 while retaining strong temporal performance. The full objective achieves the best duration accuracy and SER, although its pause F1 remains below that of pause-only optimization.

\begin{table*}[htb]
  \centering
  \small
  \caption{Reward-component ablation under a common initialization and optimization budget.}
  \label{tab:reward-ablation}
  \begin{adjustbox}{max width=\textwidth}
  \begin{tabular}{lccccccc}
    \toprule
    Reward & COMET$\uparrow$ & MetricX-QE$\downarrow$ & $\durat{2\%}\uparrow$ & $\durat{5\%}\uparrow$ & Pause F1$\uparrow$ & tIoU$\uparrow$ & SER$\downarrow$ \\
    \midrule
    Quality only & \textbf{0.795} & \textbf{3.90} & 10.2 & 38.2 & 0.354 & 0.206 & 0.134 \\
    Duration only & 0.771 & 4.40 & 59.0 & 89.2 & 0.556 & 0.316 & 0.175 \\
    
    Pause only & 0.769 & 4.47 & 54.4 & 87.6 & \textbf{0.567} & \textbf{0.334} & 0.134 \\
    Duration + Pause & 0.772 & 4.37 & 61.1 & 90.8 & 0.544 & 0.315 & 0.131\\
    \midrule
    Full (Quality + Duration + Pause) & 0.788 & 4.12 & \textbf{61.5} & \textbf{92.2} & 0.550 & 0.322 & \textbf{0.128} \\
    \bottomrule
  \end{tabular}
  \end{adjustbox}
\end{table*}

\section{Conclusion}
\label{sec:conclusion}
We presented \model, an end-to-end autoregressive S2ST system that learns to translate on the source timeline, rather than merely match utterance duration. A 642K-pair source-aligned corpus and progressive training enable the model to learn utterance-endpoint fidelity and salient pause alignment without an inference-time frame-level timing controller or post-hoc temporal correction. On English-to-Chinese translation, \model raises Dur@2\% from 25.6\% to 61.5\% and pause F1 from 0.118 to 0.550 over UniSS, with comparable COMET. Our analyses show that aligned SFT establishes timeline following, while verifier-guided RL refines timing and improves translation quality. Closing the translation-quality and naturalness gap to the cascaded baseline remains an important next step. Together, these findings support treating source timing as part of the translation target itself: learning not only what to say, but when to speak and when to remain silent.

\clearpage

\section*{Reproducibility Statement}
Data construction, reward definitions, training settings, and evaluation protocols are described in Appendices~\ref{app:data-pipeline}--\ref{app:human_eval}. We will release the training data, model checkpoints, and code upon publication, subject to applicable data licenses.

\section*{Ethics Statement}
Our work targets speech translation and dubbing.
Speaker-preserving generation could be misused for impersonation, while translation errors could misrepresent speakers.
Downstream use should involve appropriate speaker authorization and disclosure of synthetic speech, and data redistribution should respect source licenses and privacy constraints.

\section*{AI Use Statement}
We used Qwen3-ASR-1.7B, Hy-MT2-30B-A3B, and VoxCPM2 for transcription, translation, and speech synthesis during data construction. Gemini 3.1 Pro generated Chinese references for the GigaSpeech evaluation subset, which were manually reviewed.
ChatGPT assisted with drafting parts of the manuscript. The authors take full responsibility for the final paper and all AI-assisted content.


\bibliographystyle{iclr2027_conference}
\bibliography{iclr2027_conference}

@inproceedings{jia2019translatotron,
  author       = {Jia, Ye and Weiss, Ron J. and Biadsy, Fadi and Macherey, Wolfgang and Johnson, Melvin and Chen, Zhifeng and Wu, Yonghui},
  title        = {Direct speech-to-speech translation with a sequence-to-sequence model},
  booktitle    = {Proceedings of Interspeech 2019},
  pages        = {1123--1127},
  year         = {2019},
  publisher    = {ISCA},
}

@inproceedings{lee2022direct,
  author       = {Lee, Ann and Chen, Peng-Jen and Wang, Changhan and Gu, Jiatao and Popuri, Sravya and Ma, Xutai and Polyak, Adam and Adi, Yossi and He, Qing and Tang, Yun and others},
  title        = {Direct speech-to-speech translation with discrete units},
  booktitle    = {Proceedings of the 60th Annual Meeting of the Association for Computational Linguistics (Volume 1: Long Papers)},
  pages        = {3327--3339},
  year         = {2022},
  publisher    = {Association for Computational Linguistics},
}

@article{barrault2023seamlessm4t,
  author       = {Barrault, Lo{\"\i}c and Chung, Yu-An and Meglioli, Mariano Coria and Dale, David and Dong, Ning and Duquenne, Paul-Ambroise and Elsahar, Hady and Gong, Hongyu and Heffernan, Kevin and Hoffman, John and others},
  title        = {{SeamlessM4T}: Massively multilingual \& multimodal machine translation},
  journal      = {arXiv preprint arXiv:2308.11596},
  year         = {2023},
}

@article{seamless2023expressive,
  author       = {Barrault, Lo{\"\i}c and Chung, Yu-An and Meglioli, Mariano Coria and Dale, David and Dong, Ning and Duppenthaler, Mark and Duquenne, Paul-Ambroise and Ellis, Brian and Elsahar, Hady and Haaheim, Justin and others},
  title        = {{Seamless}: Multilingual expressive and streaming speech translation},
  journal      = {arXiv preprint arXiv:2312.05187},
  year         = {2023},
}

@inproceedings{cheng2025uniss,
  author       = {Cheng, Sitong and Bian, Weizhen and Wang, Xinsheng and Yuan, Ruibin and Chen, Jianyi and Yin, Shunshun and Guo, Yike and Xue, Wei},
  title        = {{UniSS}: Unified expressive speech-to-speech translation with your voice},
  booktitle    = {International Conference on Learning Representations},
  year         = {2026},
}

@inproceedings{le2024transvip,
  author       = {Le, Chenyang and Qian, Yao and Wang, Dongmei and Zhou, Long and Liu, Shujie and Wang, Xiaofei and Yousefi, Midia and Qian, Yanmin and Li, Jinyu and Zhao, Sheng and others},
  title        = {{TransVIP}: Speech to speech translation system with voice and isochrony preservation},
  booktitle    = {Advances in Neural Information Processing Systems},
  volume       = {37},
  pages        = {89682--89705},
  year         = {2024},
}

@inproceedings{tam2022isochrony,
  author       = {Tam, Derek and Lakew, Surafel M. and Virkar, Yogesh and Mathur, Prashant and Federico, Marcello},
  title        = {Isochrony-aware neural machine translation for automatic dubbing},
  booktitle    = {Proceedings of Interspeech 2022},
  pages        = {1776--1780},
  year         = {2022},
  publisher    = {ISCA},
}

@inproceedings{lakew2022isometric,
  author       = {Lakew, Surafel M. and Virkar, Yogesh and Mathur, Prashant and Federico, Marcello},
  title        = {Isometric {MT}: Neural machine translation for automatic dubbing},
  booktitle    = {Proceedings of the IEEE International Conference on Acoustics, Speech and Signal Processing (ICASSP)},
  pages        = {6242--6246},
  year         = {2022},
  publisher    = {IEEE},
}

@inproceedings{won2025duration,
  author       = {Won, Hyun-Sik and Jeong, DongJin and Choi, Hyunkyu and Kim, Jinwon},
  title        = {End-to-end multilingual automatic dubbing via duration-based translation with large language models},
  booktitle    = {Proceedings of the 2025 Conference on Empirical Methods in Natural Language Processing: System Demonstrations},
  pages        = {515--521},
  year         = {2025},
  publisher    = {Association for Computational Linguistics},
}

@article{zheng2025gspo,
  author       = {Zheng, Chujie and Liu, Shixuan and Li, Mingze and Chen, Xiong-Hui and Yu, Bowen and Gao, Chang and Dang, Kai and Liu, Yuqiong and Men, Rui and Yang, An and Zhou, Jingren and Lin, Junyang},
  title        = {Group sequence policy optimization},
  journal      = {arXiv preprint arXiv:2507.18071},
  year         = {2025},
}

@article{qwen2026asr,
  author       = {Shi, Xian and Wang, Xiong and Guo, Zhifang and Wang, Yongqi and Zhang, Pei and Zhang, Xinyu and Guo, Zishan and Hao, Hongkun and Xi, Yu and Yang, Baosong and others},
  title        = {{Qwen3-ASR} technical report},
  journal      = {arXiv preprint arXiv:2601.21337},
  year         = {2026},
}

@inproceedings{rei2020comet,
  author       = {Rei, Ricardo and Stewart, Craig and Farinha, Ana C. and Lavie, Alon},
  title        = {{COMET}: A neural framework for {MT} evaluation},
  booktitle    = {Proceedings of the 2020 Conference on Empirical Methods in Natural Language Processing (EMNLP)},
  pages        = {2685--2702},
  year         = {2020},
  publisher    = {Association for Computational Linguistics},
}

@inproceedings{wang2021covost2,
  author       = {Wang, Changhan and Wu, Anne and Gu, Jiatao and Pino, Juan},
  title        = {{CoVoST} 2 and massively multilingual speech translation},
  booktitle    = {Proceedings of Interspeech 2021},
  pages        = {2247--2251},
  year         = {2021},
  publisher    = {ISCA},
}

@inproceedings{chen2021gigaspeech,
  author       = {Chen, Guoguo and Chai, Shuzhou and Wang, Guan-Bo and Du, Jiayu and Zhang, Wei-Qiang and Weng, Chao and Su, Dan and Povey, Daniel and Trmal, Jan and Zhang, Junbo and others},
  title        = {{GigaSpeech}: An evolving, multi-domain {ASR} corpus with 10,000 hours of transcribed audio},
  booktitle    = {Proceedings of Interspeech 2021},
  pages        = {3670--3674},
  year         = {2021},
  publisher    = {ISCA},
}

@inproceedings{he2024emilia,
  author       = {He, Haorui and Shang, Zengqiang and Wang, Chaoren and Li, Xuyuan and Gu, Yicheng and Hua, Hua and Liu, Liwei and Yang, Chen and Li, Jiaqi and Shi, Peiyang and others},
  title        = {{Emilia}: An extensive, multilingual, and diverse speech dataset for large-scale speech generation},
  booktitle    = {Proceedings of the IEEE Spoken Language Technology Workshop (SLT)},
  pages        = {885--890},
  year         = {2024},
  publisher    = {IEEE},
}

@article{voxcpm2technical,
  author       = {Zhou, Yixuan and Zeng, Guoyang and Liu, Xin and Li, Xiang and Yu, Renjie and Gui, Jiancheng and Wu, Jiaheng and Wang, Ziyang and Shen, Xudong and Ye, Runchuan and others},
  title        = {{VoxCPM2} technical report},
  journal      = {arXiv preprint arXiv:2606.06928},
  year         = {2026},
}

@article{jin2025Omni,
  author       = {Xu, Jin and Guo, Zhifang and He, Jinzheng and Hu, Hangrui and He, Ting and Bai, Shuai and Chen, Keqin and Wang, Jialin and Fan, Yang and Dang, Kai and Zhang, Bin and Wang, Xiong and Chu, Yunfei and Lin, Junyang},
  title        = {{Qwen2.5-Omni} technical report},
  journal      = {arXiv preprint arXiv:2503.20215},
  year         = {2025},
}

@article{zheng2026hymt2,
  author       = {Zheng, Mao and Li, Zheng and Chen, Tao and Lv, Bo and Sun, Mingrui and Song, Mingyang and Song, Jinlong and Huang, Hong and Wu, Decheng and Wang, Hai and others},
  title        = {{Hy-MT2}: A family of fast, efficient and powerful multilingual translation models in the wild},
  journal      = {arXiv preprint arXiv:2605.22064},
  year         = {2026},
}

@inproceedings{juraska-etal-2024-metricx,
    title = "{M}etric{X}-24: The {G}oogle Submission to the {WMT} 2024 Metrics Shared Task",
    author = "Juraska, Juraj  and
      Deutsch, Daniel  and
      Finkelstein, Mara  and
      Freitag, Markus",
    editor = "Haddow, Barry  and
      Kocmi, Tom  and
      Koehn, Philipp  and
      Monz, Christof",
    booktitle = "Proceedings of the Ninth Conference on Machine Translation",
    month = nov,
    year = "2024",
}

@inproceedings{chen2025f5,
  title={F5-tts: A fairytaler that fakes fluent and faithful speech with flow matching},
  author={Chen, Yushen and Niu, Zhikang and Ma, Ziyang and Deng, Keqi and Wang, Chunhui and Zhao, Jian and Yu, Kai and Chen, Xie},
  booktitle={Proceedings of the 63rd Annual Meeting of the Association for Computational Linguistics (Volume 1: Long Papers)},
  pages={6255--6271},
  year={2025}
}

@inproceedings{gao2022paraformer,
  author       = {Gao, Zhifu and Zhang, Shiliang and McLoughlin, Ian and Yan, Zhijie},
  title        = {{Paraformer}: Fast and accurate parallel transformer for non-autoregressive end-to-end speech recognition},
  booktitle    = {Proceedings of Interspeech 2022},
  pages        = {2063--2067},
  year         = {2022},
  publisher    = {ISCA},
}

@article{du2025cosyvoice,
  author       = {Du, Zhihao and Gao, Changfeng and Wang, Yuxuan and Yu, Fan and Zhao, Tianyu and Wang, Hao and Lv, Xiang and Wang, Hui and Ni, Chongjia and Shi, Xian and others},
  title        = {{CosyVoice} 3: Towards in-the-wild speech generation via scaling-up and post-training},
  journal      = {arXiv preprint arXiv:2505.17589},
  year         = {2025},
}

@inproceedings{wu2025towards,
  author       = {Wu, Haibin and Hu, Yuxuan and Fan, Ruchao and Wang, Xiaofei and Kumatani, Kenichi and Ren, Bo and Yu, Jianwei and Lu, Heng and Wang, Lijuan and Qian, Yao and others},
  title        = {Towards efficient speech-text jointly decoding within one speech language model},
  booktitle    = {Proceedings of the IEEE Automatic Speech Recognition and Understanding Workshop (ASRU)},
  pages        = {1--7},
  year         = {2025},
  publisher    = {IEEE},
}

@misc{team2021silero,
  author       = {{Silero Team}},
  title        = {{Silero VAD}: Pre-trained enterprise-grade voice activity detector ({VAD}), number detector and language classifier},
  year         = {2021},
  howpublished = {GitHub repository},
}

@article{chen2022wavlm,
  author       = {Chen, Sanyuan and Wang, Chengyi and Chen, Zhengyang and Wu, Yu and Liu, Shujie and Chen, Zhuo and Li, Jinyu and Kanda, Naoyuki and Yoshioka, Takuya and Xiao, Xiong and others},
  title        = {{WavLM}: Large-scale self-supervised pre-training for full stack speech processing},
  journal      = {IEEE Journal of Selected Topics in Signal Processing},
  volume       = {16},
  number       = {6},
  pages        = {1505--1518},
  year         = {2022},
  publisher    = {IEEE},
}

@article{steinhauer1999brain,
  author       = {Steinhauer, Karsten and Alter, Kai and Friederici, Angela D.},
  title        = {Brain potentials indicate immediate use of prosodic cues in natural speech processing},
  journal      = {Nature Neuroscience},
  volume       = {2},
  number       = {2},
  pages        = {191--196},
  year         = {1999},
}

@inproceedings{virkar2022prosodic,
  author       = {Virkar, Yogesh and Federico, Marcello and Enyedi, Robert and Barra-Chicote, Roberto},
  title        = {Prosodic alignment for off-screen automatic dubbing},
  booktitle    = {Proceedings of Interspeech 2022},
  pages        = {496--500},
  year         = {2022},
  publisher    = {ISCA},
}

@article{brannon2023dubbing,
  author       = {Brannon, William and Virkar, Yogesh and Thompson, Brian},
  title        = {Dubbing in practice: A large scale study of human localization with insights for automatic dubbing},
  journal      = {Transactions of the Association for Computational Linguistics},
  volume       = {11},
  pages        = {419--435},
  year         = {2023},
}


\clearpage
\appendix

\section{Extended Related Work}

\paragraph{Direct and expressive speech-to-speech translation.}
Early direct S2ST systems demonstrated speech translation through spectrogram prediction~\citep{jia2019translatotron} and discrete-unit generation~\citep{lee2022direct}, while SeamlessM4T extended unified speech and text translation to a massively multilingual setting~\citep{barrault2023seamlessm4t}. Beyond semantic accuracy, SeamlessExpressive models vocal style and prosody, including speech rate and pauses~\citep{seamless2023expressive}, and UniSS combines speech semantics and style within a unified text--speech language model to preserve speaker identity, emotion, and duration consistency~\citep{cheng2025uniss}. These developments establish expressive transfer as an important dimension of S2ST. Our work focuses on a complementary requirement: preserving the source's speech--silence structure on a shared absolute timeline. Preserving voice or speaking rate does not guarantee duration agreement. Even matching total duration does not ensure that internal pauses occur at corresponding moments.

\paragraph{Isochrony and automatic dubbing.}
Temporal constraints have been studied at both the translation and speech-generation levels. Isometric MT encourages translations to match the source text length~\citep{lakew2022isometric}, while duration-based translation uses source-speech duration to determine phoneme budgets and iteratively adjust translated text~\citep{won2025duration}. Importantly, prior work also addresses internal structure: isochrony-aware MT jointly translates and projects pause markers~\citep{tam2022isochrony}, and TransVIP conditions speech generation on duration and frame-level voice activity through an explicit isochrony-control module~\citep{le2024transvip}. Whereas TransVIP derives its isochrony conditioning from target speech during training and source speech at inference, we construct target waveforms that already follow the source timeline. This supervision enables \model to learn endpoint fidelity, salient pause alignment, and autonomous termination jointly, without post-hoc padding, truncation, or global time stretching at inference.

\paragraph{Speech timing and context-dependent synchronization.}
Psycholinguistic evidence indicates that intonational phrasing contributes to the initial analysis of sentence structure and that misleading prosodic boundaries can disrupt interpretation~\citep{steinhauer1999brain}. These findings concern prosodic structure, rather than exact cross-lingual timestamp correspondence. Dubbing research further suggests that synchronization should be considered together with translation quality, vocal naturalness, and visual context. \citet{brannon2023dubbing} analyze professional human dubs and argue for a qualified view of timing constraints. \citet{virkar2022prosodic} distinguish on-screen from off-screen dubbing and report improved subjective viewing experience when the alignment strategy accounts for speaker visibility. Against this background, we study how to learn source-timeline following when that constraint is required, rather than prescribing a single synchronization policy for all translation scenarios.

\paragraph{Joint text--speech modeling and reward-based post-training.}
Joint speech language models provide a common sequence space for text and speech generation, with interleaved decoding coordinating the two modalities~\citep{wu2025towards}. Reward-based post-training has also been explored in speech synthesis, with CosyVoice~3 improving content consistency beyond standard likelihood training~\citep{du2025cosyvoice}. For language-model optimization, GSPO uses sequence-level importance ratios and clipping to align policy updates with sequence-level rewards~\citep{zheng2025gspo}. We adapt these ingredients to timeline-preserving S2ST by separating semantic and temporal feedback within the interleaved output. Translation quality, global duration agreement, and local pause alignment are normalized independently, with semantic feedback assigned to translation-text tokens and temporal feedback to speech-codec tokens. 



\section{Data Construction and Filtering Details}
\label{app:data-pipeline}

This appendix complements Section~\ref{sec:data} with the source-selection criteria, construction settings, and quality checks. We follow the same three-stage organization: Stage~2 synthesizes the natural Chinese translation \(z^{\mathrm{nat}}\), whereas Stage~3 generates new translations and speech from the Stage~1 temporal scaffold and English region transcripts \(e_{1:N}\). Stage~3 does not reuse \(z^{\mathrm{nat}}\) or the Stage~2 target waveform. Figure~\ref{fig:data_stage3_canvas} illustrates the Stage~3 construction.

\subsection{Source selection}
\label{app:source-selection}

We draw source utterances from the training partitions of Emilia-EN~\citep{he2024emilia} and GigaSpeech~\citep{chen2021gigaspeech}. Stages~1 and~2 use utterances of 8--25\,s with at least eight words. Stage~3 additionally includes a short-form pool of 2--8\,s with at least three words. Short-form sources undergo the same source-side processing but contribute only to Stage~3.

Both pools require an estimated speaking rate of 3.0--5.5 syllables/s and leading and trailing non-speech intervals no longer than 1.5\,s each. Within Stage~3, no-pause examples are restricted to 2--12\,s, whereas pause-aligned examples may extend to 25\,s.

\subsection{Stage 1: temporal scaffold and text grounding}
\label{app:pause-scaffold}

\paragraph{Source-derived speech regions and pauses.}
We use Silero VAD~\citep{team2021silero} to identify source speech spans and the internal gaps between them. Gaps of at least 0.5\,s define the selected pause boundaries. Speech regions are split only at these gaps, with the following region anchored to the next strict-VAD speech onset. Leading and trailing non-speech remain part of the source canvas but are not internal pause labels. The strict pass uses a speech-posterior threshold of 0.8, minimum speech and silence durations of 250 and 100\,ms, and 30-ms boundary padding. A separate 0.4-s threshold is used for diagnostic gap detection, not for selecting supervised pauses.

\paragraph{Source-speech leakage screening.}
Because source-canvas rendering retains audio outside the speech windows, we screen these intervals for source speech missed by the strict VAD pass. A second, more permissive pass with a speech-posterior threshold of 0.35 flags speech spans with less than 90\% coverage by strict-VAD spans expanded by 0.15\,s on each side. Utterances with a suspected leakage region of at least 0.6\,s are discarded. Remaining suspect regions of at least 0.1\,s are checked with Qwen3-ASR-1.7B~\citep{qwen2026asr}; any nonempty hypothesis triggers rejection.

\paragraph{Region transcription.}
The same ASR model transcribes the source speech regions to obtain \(e_{1:N}\). All required transcripts must be nonempty, and their number-normalized word error rate against the corpus transcript must not exceed 0.15. We join the region transcripts with \texttt{<pause>} markers only at the selected source boundaries, yielding the Stage~1 ASR target described in Section~\ref{sec:data}.

\paragraph{Natural translation.}
Hy-MT2-30B-A3B~\citep{zheng2026hymt2} translates the complete utterance into \(z^{\mathrm{nat}}\). The prompt requests faithful, colloquial Chinese without unnecessary fillers and imposes no duration, character-count, or pause-placement constraint. If the output contains Latin letters, one additional request explicitly prohibits them; the retry replaces the original only when it is entirely Latin-free. This translation provides the Stage~1 AST target and the Stage~2 synthesis text.

\subsection{Stage 2: natural-pace speech supervision}
\label{app:natural-synthesis}

VoxCPM2~\citep{voxcpm2technical} synthesizes \(z^{\mathrm{nat}}\) in a single whole-utterance call conditioned on source-speaker reference audio. The target follows the natural phrasing and duration of the Chinese translation, without segment chaining, synthesis-guided retranslation, time-scale modification, or source-canvas rendering. The acceptance criteria in Table~\ref{tab:data-quality-checks} remove unreliable synthesis without imposing source-timeline alignment. In particular, the broad target-to-source duration-ratio range is a sanity check rather than an isochrony constraint.

\subsection{Stage 3: timeline-preserving speech supervision}
\label{app:iso-synthesis}

We form the no-pause and pause-aligned subsets from the shared source scaffold. No-pause examples contain no selected internal pause, although shorter hesitations may remain within a speech region. Pause-aligned examples contain one to five selected pauses, each approximately 0.5--1.5\,s long, with every speech region lasting at least 0.8\,s. All selected pauses are retained.

Let \([a_i,b_i]\) denote source speech region \(i\), with available speaking time \(d_i=b_i-a_i\), measured in seconds. Stage~3 generates a new Chinese translation and speech segment for each region while keeping these windows fixed.

\paragraph{Duration-budgeted translation.}
\label{app:budgeted-translation}
We initialize the character budget from the source-region duration using a reference rate of 4.8 Chinese characters per second:
\begin{equation}
    m_i = 4.8d_i, \qquad
    L_i = \max\!\left(2,\left\lceil0.85m_i\right\rceil\right),\qquad
    U_i = \max\!\left(3,\left\lceil1.15m_i\right\rceil\right).
    \label{eq:implemented-budget}
\end{equation}
Only Chinese ideographs count toward the budget; punctuation and whitespace are excluded. The translation prompt provides the complete transcript \(e_{1:N}\) as semantic context, the current region index and transcript \(e_i\), and the requested character range \([L_i,U_i]\).

We generate up to three fixed-seed translation candidates. Candidates within the requested range are preferred to those outside it. Within either group, candidates closer to the interval center are preferred. The selected text undergoes the same Latin-removal retry as the natural translation.

If the selected translation cannot be synthesized within the permitted temporal range, the speaking rate of the closest synthesis candidate is used to tighten the character budget and retranslate the same English region. This feedback changes the translated text, not the source speech windows or pause boundaries.

\paragraph{Speaker-conditioned synthesis.}
VoxCPM2 generates candidates for each translated region using the longest detected source speech span, capped at 10\,s, as a common speaker reference. For each region after the first, synthesis additionally conditions on the preceding selected target segment and its text. This preceding segment is used before duration correction to encourage local continuity. Each region receives at most two synthesis rounds, with six fixed-seed candidates per round.

\paragraph{Candidate acceptance and selection.}
After VAD-based edge trimming, let \(q_{i,j}\) and \(s_{i,j}\) denote the duration and speaker similarity of candidate \(j\) for region \(i\). Ordinary candidates are eligible for duration correction only when
\[
    \frac{|q_{i,j}-d_i|}{d_i}\leq 0.10
    \qquad\text{and}\qquad
    s_{i,j}\geq 0.45.
\]
A separate short-tail option accepts candidates satisfying \(0\leq d_i-q_{i,j}\leq0.15\,\mathrm{s}\) under the same speaker-similarity threshold. These candidates retain their natural duration rather than being stretched.

Among eligible candidates, selection jointly favors duration agreement, a duration ratio \(q_{i,j}/d_i\) close to that of the preceding selected segment, higher speaker similarity, and fewer generated internal gaps of at least 0.4\,s. Internal gaps incur a soft penalty rather than causing automatic rejection.

Generation may stop early when a candidate achieves at most 2\% duration error, speaker similarity of at least 0.60, and no detected internal gap. These are early-stopping conditions, not requirements for every accepted candidate. If even one region has no acceptable candidate, the complete utterance is discarded.

\paragraph{Duration correction and source-canvas rendering.}
\label{app:rendering}
We fit each accepted target segment to its assigned source window. For an ordinary selected candidate of duration \(q_i\), Rubber Band R3 applies the factor \(d_i/q_i\) using the finer engine with formant preservation. The corrected segment is then cropped or padded to the exact window sample count. Short-tail candidates are not stretched; source-derived room tone fills the remaining window. Candidates outside the acceptance ranges are rejected rather than subjected to larger corrections.

We construct the target waveform by replacing the speech in each source window \([a_i,b_i]\) with the corresponding target segment while retaining the source audio outside these windows. Source-derived room tone provides background ambience beneath the replacement speech. We match the source-region RMS level with gain limited to \(\pm6\)\,dB and apply 20-ms equal-power cross-fades at replacement boundaries.

The completed waveform is stored as 16-kHz PCM16 with exactly the same sample count as the source, keeping the selected source pause intervals and the utterance endpoint fixed on the source canvas. These operations construct training targets and are not applied during model inference.

\subsection{Quality control and retained data}
\label{app:quality-filtering}

\paragraph{Acceptance criteria.}
Table~\ref{tab:data-quality-checks} summarizes the target-speech acceptance criteria. Chinese-character error rate (CER) compares Paraformer-zh re-ASR~\citep{gao2022paraformer} with the intended synthesis text after retaining only Chinese ideographs. Stage~3 translation quality is evaluated with MetricX-24 QE (XXL), using the concatenated English region transcripts and intended Chinese translations. Weak-speech screening takes place before translation, and Stage~3 speaker similarity is enforced during candidate selection rather than checked again at final verification.

\begin{table}[t]
    \centering
    \caption{Acceptance criteria for synthesized training targets. Stage~3 speaker checks are applied during candidate selection.}
    \label{tab:data-quality-checks}
    \small
    \setlength{\tabcolsep}{5pt}
    \renewcommand{\arraystretch}{1.12}
    \begin{tabularx}{\linewidth}{@{}p{0.22\linewidth}>{\raggedright\arraybackslash}X>{\raggedright\arraybackslash}X@{}}
        \toprule
        Criterion & Stage~2: natural pace & Stage~3: timeline preserving \\
        \midrule
        Content realization & Chinese CER \(\leq0.10\) & Chinese CER \(\leq0.10\) \\
        Script purity & No Latin letters in re-ASR & No Latin letters in intended text or re-ASR \\
        Speaker consistency & Similarity \(\geq0.45\) & Similarity \(\geq0.45\) per selected segment \\
        Translation adequacy & No QE gate & MetricX-24 QE \(\leq5.0\) \\
        Duration & Target/source ratio in \([0.35,2.0]\) & Exact source sample count \\
        \bottomrule
    \end{tabularx}
\end{table}

\paragraph{Timing audits.}
Pause tIoU and source-pause matching with a 0.4-s tolerance are computed for auditing, not used as hard acceptance gates. Endpoint equality follows from exact sample-count matching, while the selected source pause intervals are retained through source-canvas rendering.

\begin{figure}[htb]
    \centering
    \begin{subfigure}[t]{0.48\linewidth}
        \centering
        \includegraphics[width=\linewidth]{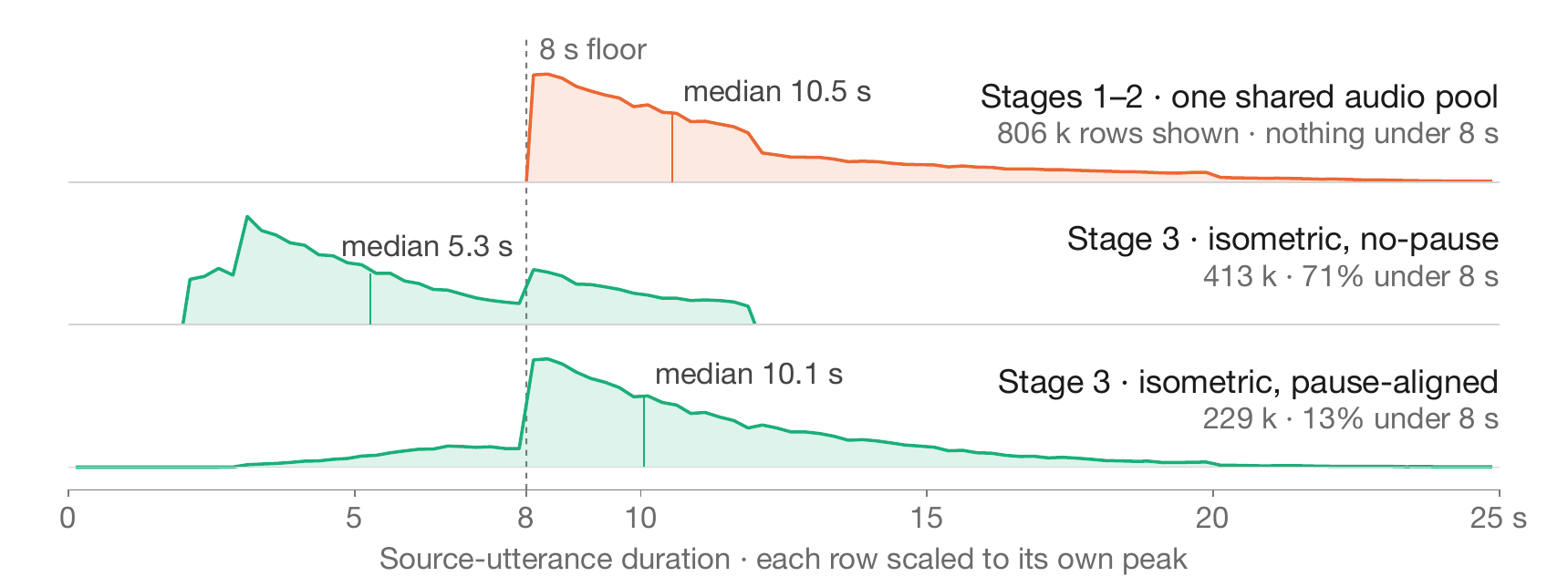}
        \caption{Source data duration distributions.}
        \label{fig:data-source-durations}
    \end{subfigure}
    \hfill
    \begin{subfigure}[t]{0.5\linewidth}
        \centering
        \includegraphics[width=\linewidth]{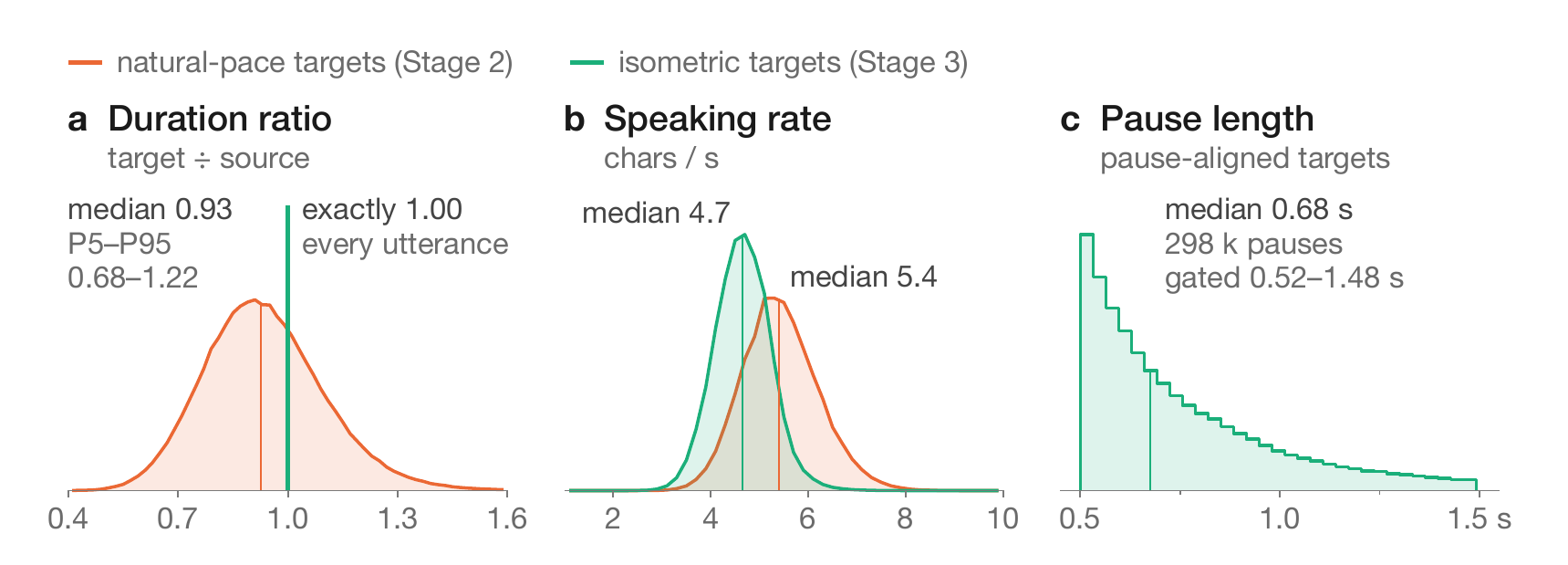}
        \caption{Target timing statistics.}
        \label{fig:data-target-timing}
    \end{subfigure}

    \caption{Training-data timing statistics. (a) Source-duration distributions, including short utterances added in Stage 3. (b) Target/source duration ratios, speaking rates, and supervised pause lengths. Stage 3 targets exactly match source duration, unlike Stage 2 natural-pace targets.}
    \label{fig:data}
\end{figure}

\paragraph{Data export and retained corpus.}
Verified waveforms are encoded with s3tokenizer-v3 at 25 tokens/s using its 6,561-entry codebook. Training records retain all supervised pause spans and waveform timing metadata. These annotations are consumed directly without rerunning VAD or deriving pauses from punctuation.

Table~\ref{tab:training-data} and Figs.~\ref{fig:data-source-durations} and~\ref{fig:data-target-timing} summarize the retained data.
Stage~3 contains 642,276 pairs and 1,368.8 hours per side, including 320,421 short-form examples below 8\,s and 321,855 examples of at least 8\,s. Reported hours count training exposure across tasks and stages, not unique source audio.

\section{Reward definitions and policy-optimization details}
\label{app:reward-details}

This appendix details the verifier rewards and policy update used in Stage~4 (Section~\ref{sec:rl}). We first specify the reward functions, then describe component-wise advantage estimation and masked policy optimization. The temporal rewards are illustrated in Figure~\ref{fig:reward}. All temporal quantities are measured in seconds.

\subsection{Translation-quality and global-duration rewards}
\label{app:quality-duration-rewards}

\paragraph{Translation quality.}
We use the reference-free MetricX-QE reward,

\begin{equation}
  R_{\mathrm{qual}}(x,y)
  =
  1 - \frac{
    \operatorname{MetricX\text{-}QE}(s_x,\hat{t}_y)
  }{25}.
  \label{eq:quality-reward}
\end{equation}

The verifier compares the generated translation text $\hat{t}_y$ with the dataset-provided source transcript $s_x$, rather than the model's own ASR output. The linear transformation maps the lower-is-better MetricX-QE score in $[0,25]$ to a higher-is-better reward in $[0,1]$.

\paragraph{Global duration.}
Let $T_x$ and $T_y$ denote the source and generated speech durations, respectively, and let $\rho=T_y/T_x$. We allow a free region of $\epsilon_{\mathrm{free}}=0.02$ and use $\epsilon_{\mathrm{ref}}=0.10$ to calibrate the penalty outside it. The normalized log-distance is
\begin{equation}
  u(\rho)=
  \begin{cases}
    0,
    & |\rho-1|\leq\epsilon_{\mathrm{free}},
    \\[5pt]
    \dfrac{\log\!\left(\rho/(1+\epsilon_{\mathrm{free}})\right)}
    {\log\!\left((1+\epsilon_{\mathrm{ref}})
                 /(1+\epsilon_{\mathrm{free}})\right)},
    & \rho>1+\epsilon_{\mathrm{free}},
    \\[12pt]
    \dfrac{\log\!\left((1-\epsilon_{\mathrm{free}})/\rho\right)}
    {\log\!\left((1-\epsilon_{\mathrm{free}})
                 /(1-\epsilon_{\mathrm{ref}})\right)},
    & \rho<1-\epsilon_{\mathrm{free}}.
  \end{cases}
  \label{eq:app-duration-distance}
\end{equation}
The corresponding reward is
\begin{equation}
  R_{\mathrm{dur}}(x,y)
  =-\min\!\left(u(\rho)^2,\,4\right).
  \label{eq:app-duration-reward}
\end{equation}
Thus, the reward is zero within $\pm2\%$ of the source duration, equals $-1$ at $\rho=0.9$ and $\rho=1.1$, and saturates at $-4$ for large mismatches.

\subsection{Timeline-based pause reward}
\label{app:pause-reward-details}

The pause reward compares internal silence on a shared absolute timeline. It combines missed and spurious silence, pause-center displacement, and utterance-end mismatch.

\paragraph{Pause preprocessing.}
The temporal origin is the beginning of each utterance. Leading and trailing silence is excluded from the pause-event sets, without shifting the timestamps of internal pauses. Silence intervals shorter than $\tau_{\mathrm{sil}}=0.12$\,s are discarded. For generated speech, neighboring silence intervals separated by at most two codec tokens are merged; at 25 codec tokens per second, this gives $\tau_{\mathrm{merge}}=2/25=0.08$\,s. After preprocessing, the source and generated pause sets are
\[
  \mathcal{I}_x=\{[a_k^x,b_k^x]\}_{k=1}^{K_x},
  \qquad
  \mathcal{I}_y=\{[a_j^y,b_j^y]\}_{j=1}^{K_y},
\]
where $K_x$ and $K_y$ are their respective event counts. The generated-silence occupancy is $\mathcal{S}_y=\bigcup_{j=1}^{K_y}[a_j^y,b_j^y]$.

\paragraph{Collar-tolerant occupancy errors.}
A boundary collar of $\delta=0.1$\,s tolerates small differences in pause boundaries. We erode each source pause to obtain a silence core and dilate it to obtain a tolerated silence region:
\begin{align}
  \mathcal{S}_x^{\mathrm{core}}
  &=\bigcup_{\substack{1\leq k\leq K_x\\ b_k^x-a_k^x>2\delta}}
    [a_k^x+\delta,\,b_k^x-\delta],
  \label{eq:app-pause-core}\\
  \mathcal{S}_x^{\mathrm{tol}}
  &=\bigcup_{k=1}^{K_x}
    \left([a_k^x-\delta,\,b_k^x+\delta]\cap[0,T_x]\right).
  \label{eq:app-pause-tolerance}
\end{align}
Intervals that become empty after erosion contribute no silence core. Let $\mu(\cdot)$ denote total interval duration. We define
\begin{align}
  m(x,y)&=\mu\!\left(
    \mathcal{S}_x^{\mathrm{core}}\setminus\mathcal{S}_y\right),
  \label{eq:app-pause-miss}\\
  f(x,y)&=\mu\!\left(
    \mathcal{S}_y\setminus\mathcal{S}_x^{\mathrm{tol}}\right),
  \label{eq:app-pause-false-alarm}\\
  o(x,y)&=|T_y-T_x|.
  \label{eq:app-pause-overhang}
\end{align}
Here, $m$ measures uncovered source-silence cores, $f$ measures generated silence outside the tolerated source regions, and $o$ measures endpoint mismatch, including both early and late termination.

\paragraph{Pause-center displacement.}
For pause centers $c_k^x=(a_k^x+b_k^x)/2$ and $c_j^y=(a_j^y+b_j^y)/2$, each source event is compared with its nearest generated event. The total displacement error is
\begin{equation}
  \begin{aligned}
    d(x,y)&=\sum_{k=1}^{K_x}\Delta_k,\\
    \Delta_k&=
    \begin{cases}
      \displaystyle
      \min\!\left(\min_{1\leq j\leq K_y}|c_k^x-c_j^y|,
                  \,\tau_{\mathrm{drift}}\right),
      & K_y>0,\\[6pt]
      \tau_{\mathrm{drift}}, & K_y=0,
    \end{cases}
  \end{aligned}
  \label{eq:app-pause-drift}
\end{equation}
with $\tau_{\mathrm{drift}}=2.0$\,s. A source event receives the maximum displacement penalty when no generated pause is present or its nearest one is more than 2.0\,s away. These nearest-event comparisons do not impose one-to-one matching. When $K_x=0$, the sum is zero.

\paragraph{Combined penalty and duration gate.}
The complete pause penalty is
\begin{align}
  P_{\mathrm{pause}}(x,y)
  ={}&\frac{
    1.5\,m(x,y)+1.5\,f(x,y)+2.0\,o(x,y)
    +\mathbf{1}[K_x>0]\,d(x,y)
  }{\max(T_x,\epsilon)}
  \nonumber\\
  &+0.3\,\mathbf{1}[K_x>0\land K_y=0],
  \label{eq:app-pause-penalty}
\end{align}
where $\epsilon>0$ safeguards duration normalization. The time-based errors are normalized by the source duration; an additional penalty of $0.3$ applies when the source contains pauses but none is generated. We first clip the negative penalty from below,
\begin{equation}
  R_{\mathrm{pause}}^{0}(x,y)
  =\max\!\left(-P_{\mathrm{pause}}(x,y),\,r_{\mathrm{floor}}\right),
  \label{eq:app-pause-reward}
\end{equation}
where $r_{\mathrm{floor}}=-4$ denotes the pre-gate reward-clipping floor used during training.
The final reward then applies the global-duration gate:
\begin{equation}
  R_{\mathrm{pause}}(x,y)=
  \begin{cases}
    \min\!\left(R_{\mathrm{pause}}^{0}(x,y),-1\right),
    & |\rho-1|>0.25,\\[3pt]
    R_{\mathrm{pause}}^{0}(x,y), & \text{otherwise}.
  \end{cases}
  \label{eq:app-pause-duration-gate}
\end{equation}
This prevents a candidate with a large duration mismatch from receiving a favorable pause reward solely through local overlap.

\subsection{Component-wise advantage estimation}
\label{app:component-advantages}

For each source prompt, we sample $G=16$ candidates. Let $R_{i,c}$ be the reward of candidate $i$ for component $c\in\{\mathrm{qual},\mathrm{pause},\mathrm{dur}\}$. Each component is standardized independently within the rollout group:
\begin{equation}
  \overline{R}_c=\frac{1}{G}\sum_{i=1}^{G}R_{i,c},
  \qquad
  Z_{i,c}=\frac{R_{i,c}-\overline{R}_c}
  {\max\!\left(\operatorname{Std}_{j=1}^{G}[R_{j,c}],\,0.05\right)}.
  \label{eq:app-component-standardization}
\end{equation}
We clip the standardized values and then recenter them:
\begin{equation}
  \widetilde{A}_{i,c}=\operatorname{clip}(Z_{i,c},-3,3),
  \qquad
  A_{i,c}=\widetilde{A}_{i,c}
          -\frac{1}{G}\sum_{j=1}^{G}\widetilde{A}_{j,c}.
  \label{eq:app-group-advantage}
\end{equation}
Clipping therefore precedes recentering, and the resulting advantages have zero mean within each component. Rewards are not combined before this calculation. Extreme-rollout filtering is disabled: all $G$ candidates are retained, and each contributes to a component when its corresponding token mask is nonempty.

\subsection{Masked GSPO and KL regularization}
\label{app:masked-policy-optimization}

\paragraph{Reward-specific token masks.}
The translation-text mask $M_{\mathrm{text}}$ selects only generated translation tokens after \texttt{<s2st\_text>} and before \texttt{<s2st\_eos>}. It excludes the control markers, speech-codec tokens, and sequence-level end-of-sequence tokens. The speech mask $M_{\mathrm{speech}}$ selects only generated speech-codec tokens. Thus,
\begin{equation}
  M_{\mathrm{qual}}=M_{\mathrm{text}},
  \qquad
  M_{\mathrm{pause}}=M_{\mathrm{dur}}=M_{\mathrm{speech}}.
  \label{eq:app-component-masks}
\end{equation}

\paragraph{Masked sequence-level policy loss.}
For candidate $y_i$ and component $c$, let $M_{i,c,t}\in\{0,1\}$ indicate whether token $t$ is selected, with mask length $L_{i,c}=\sum_t M_{i,c,t}$. For a nonempty mask, the length-normalized importance ratio of GSPO~\citep{zheng2025gspo} becomes
\begin{equation}
  s_{i,c}(\theta)=\exp\!\left(
    \frac{1}{L_{i,c}}\sum_t M_{i,c,t}
    \log\frac{
      \pi_{\theta}(y_{i,t}\mid x,y_{i,<t})
    }{
      \pi_{\theta_{\mathrm{old}}}(y_{i,t}\mid x,y_{i,<t})
    }
  \right),
  \label{eq:app-masked-sequence-ratio}
\end{equation}
where $\pi_{\theta_{\mathrm{old}}}$ is the rollout policy. The mask selects the probabilities entering the ratio; each probability remains conditioned on the full preceding sequence. We use asymmetric clipping with $\epsilon_{\mathrm{low}}=0.2$ and $\epsilon_{\mathrm{high}}=0.28$:
\[
  \overline{s}_{i,c}(\theta)
  =\operatorname{clip}\!\left(
    s_{i,c}(\theta),\,1-\epsilon_{\mathrm{low}},
    \,1+\epsilon_{\mathrm{high}}\right).
\]
Under the loss-minimization convention, the component surrogate is
\begin{equation}
  \mathcal{L}_{\mathrm{GSPO}}^{(c)}
  =-\frac{1}{G}\sum_{i=1}^{G}
    \min\!\left(
      s_{i,c}(\theta)A_{i,c},
      \,\overline{s}_{i,c}(\theta)A_{i,c}\right).
  \label{eq:app-gspo-surrogate}
\end{equation}

\paragraph{Loss aggregation and regularization.}
The component losses are combined only after their advantages and masked importance ratios have been computed:
\begin{equation}
  \mathcal{L}_{\mathrm{policy}}
  =\frac{
    \lambda_q\mathcal{L}_{\mathrm{GSPO}}^{(\mathrm{qual})}
    +\lambda_p\mathcal{L}_{\mathrm{GSPO}}^{(\mathrm{pause})}
    +\lambda_d\mathcal{L}_{\mathrm{GSPO}}^{(\mathrm{dur})}
  }{\lambda_q+\lambda_p+\lambda_d},
  \label{eq:app-component-objective}
\end{equation}
with $\lambda_q=0.2$ and $\lambda_p=\lambda_d=1$. The fixed Stage~3 checkpoint serves as the reference policy $\pi_{\mathrm{SFT}}$. We apply a single KL term over $M_{\mathrm{KL}}=M_{\mathrm{text}}\cup M_{\mathrm{speech}}$, rather than a separate term for each reward component:
\begin{equation}
  \mathcal{L}
  =\mathcal{L}_{\mathrm{policy}}
   +\beta\,D_{\mathrm{KL}}^{M_{\mathrm{KL}}}
      \!\left(\pi_{\theta}\,\|\,\pi_{\mathrm{SFT}}\right),
  \qquad \beta=0.01.
  \label{eq:app-total-objective}
\end{equation}
Both the reference policy and the KL coefficient remain fixed throughout Stage~4. With these component weights, this recovers Eq.~\ref{eq:component-gspo} in the main text.

\section{Training Hyperparameters}
\label{app:training-hyperparameters}

Table~\ref{tab:training-hyperparameters} summarizes Stages~2--4, each initialized from the final checkpoint of the preceding stage. Stage~4 samples its source prompts from the Stage~3 training set. All three stages use eight GPUs, random seed 42, and AdamW with $(\beta_1,\beta_2)=(0.9,0.999)$, $\epsilon=10^{-6}$, and gradient clipping at 1.0. Learning rates warm up over the first 5\% of updates and then decay to zero.

\begin{table}[htbp]
    \centering
    \caption{Training hyperparameters for natural-pace SFT (Stage~2), timeline-preserving SFT (Stage~3), and GSPO (Stage~4). Dashes indicate inapplicable settings.}
    \label{tab:training-hyperparameters}
    \small
    \setlength{\tabcolsep}{4pt}
    \renewcommand{\arraystretch}{1.1}
    \begin{tabularx}{\linewidth}{@{}lYYY@{}}
        \toprule
        Hyperparameter & Stage~2 & Stage~3 & Stage~4 \\
        \midrule
        Peak learning rate & $10^{-4}$ & $10^{-4}$ & $3\times10^{-6}$ \\
        Weight decay & 0.02 & 0.02 & 0.01 \\
        LR decay schedule & Linear & Linear & Cosine \\
        Optimization steps & 20,000 & 12,000 & 100 \\
        Global batch per update & 256 utterances & 256 utterances & 28 prompts \\
        Codec-parameter LR multiplier & $10\times$ & $10\times$ & -- \\
        Pause-codec loss weight & 1.0 & 2.0 & -- \\
        End-of-sequence loss weight & 4.0 & 2.0 & -- \\
        \midrule
        Rollouts per prompt & -- & -- & 16 \\
        Training microbatch size & -- & -- & 4 \\
        Rollout temperature / top-$p$ & -- & -- & 1.0 / 1.0 \\
        Maximum new tokens & -- & -- & 896 \\
        GSPO clipping $(\epsilon_{\mathrm{low}},\epsilon_{\mathrm{high}})$ & -- & -- & (0.2, 0.28) \\
        KL coefficient & -- & -- & 0.01 \\
        Component weights $(\lambda_q,\lambda_p,\lambda_d)$ & -- & -- & (0.2, 1, 1) \\
        \bottomrule
    \end{tabularx}
\end{table}

For SFT, the effective batch size is $8\times8\times4=256$ (GPUs $\times$ per-GPU batch $\times$ gradient accumulation), with BF16 mixed precision and sequence-normalized loss. Stage~4 uses seven training GPUs and one reward-server GPU; four generation batches per update yield $7\times4=28$ prompts and 448 rollouts before zero-signal groups are discarded. The policy weights are FP32, while reference weights and computation use BF16. We use the checkpoint after 100 updates as the final model.

\section{Isochrony and pause-alignment metrics}
\label{app:timing_metrics}

\begin{figure}[htb]
    \centering
    \includegraphics[width=1\linewidth]{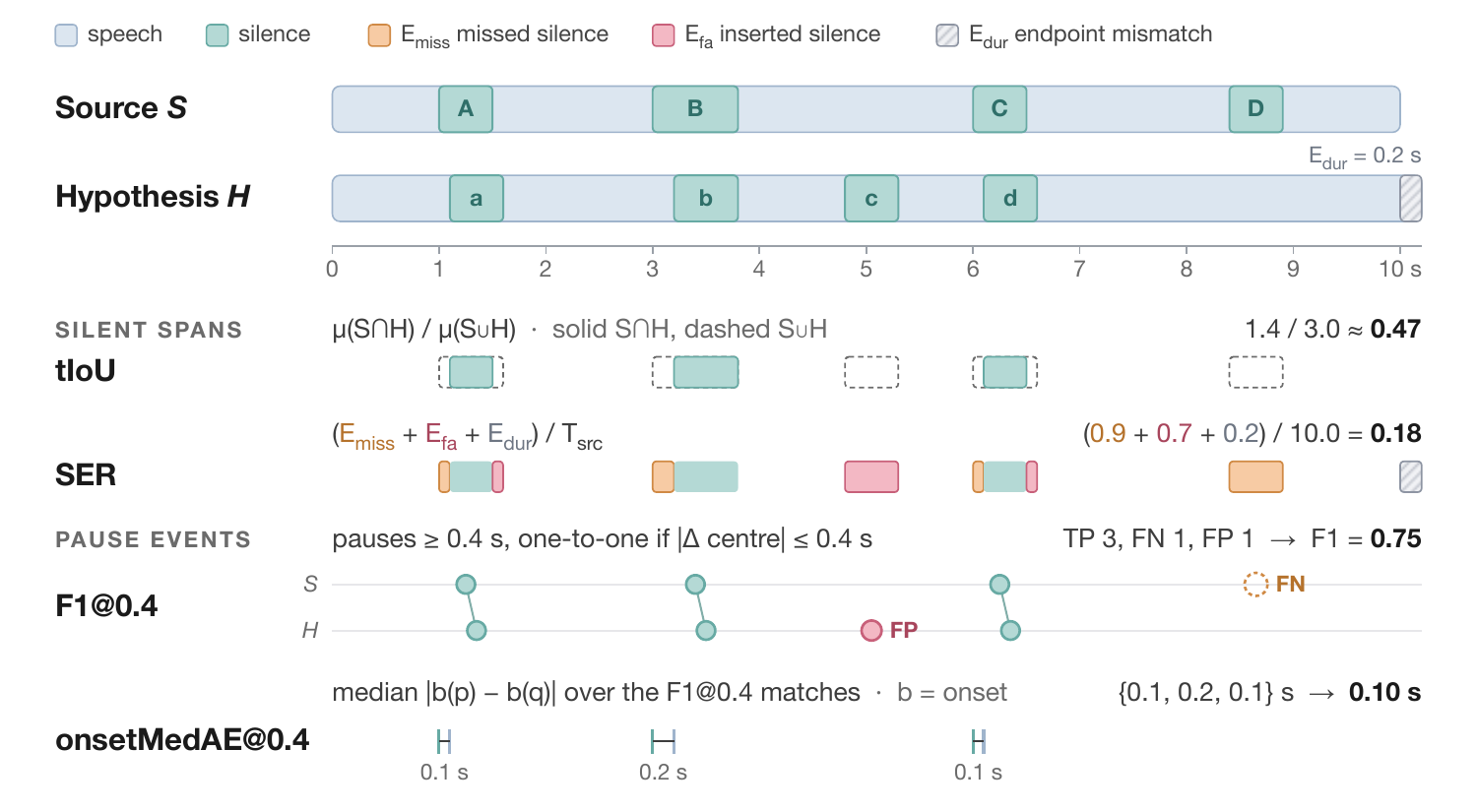}
    \caption{Pause-alignment metrics on one shared timeline. Source $S$ and hypothesis $H$ pauses (all $\geq\tau_{\min}=0.4$\,s) are compared at the same temporal origin. tIoU is the overlap of the silent spans; SER sums the missed silence $E_{\mathrm{miss}}$, the inserted silence $E_{\mathrm{fa}}$, and the endpoint mismatch $E_{\mathrm{dur}}$, normalized by the source duration. F1@0.4 matches pause events one-to-one when their centers differ by at most 0.4\,s ($D$ is a false negative, $c$ a false positive); onsetMedAE@0.4 is the median absolute onset difference over the matched pairs.
}
    \label{fig:pause_metrics}
\end{figure}

\paragraph{Isochrony.}
For utterance $i$, let $T_i^{\mathrm{src}}$ and $T_i^{\mathrm{hyp}}$ denote the durations of the source and translated speech, respectively. We define the relative duration error as
\begin{equation}
    e_i^{\mathrm{dur}}
    =
    \frac{
        \left|T_i^{\mathrm{hyp}}-T_i^{\mathrm{src}}\right|
    }{
        T_i^{\mathrm{src}}
    }.
\end{equation}
The metric $\durat{\epsilon}$ is the percentage of utterances satisfying $e_i^{\mathrm{dur}}\leq\epsilon$:
\begin{equation}
    \durat{\epsilon}
    =
    \frac{100}{N}
    \sum_{i=1}^{N}
    \mathbb{I}
    \left[
        e_i^{\mathrm{dur}}\leq\epsilon
    \right].
\end{equation}
We report $\epsilon\in\{2\%,5\%,10\%\}$; higher values indicate better utterance-level duration matching.

\paragraph{Pause alignment.}
As illustrated in Fig.~\ref{fig:pause_metrics}, we anchor the source and translated utterances at the same temporal origin and compare their pauses on a shared absolute time axis. Let $S$ and $H$ denote the unions of the source and hypothesis silent intervals, respectively, and let $\mu(\cdot)$ denote total interval duration. The metrics provide two complementary views: tIoU and SER operate on \emph{silent spans}, whereas F1 and onset error operate on discrete \emph{pause events}.

\paragraph{Pause F1.}
For the event-based view, we retain pauses of duration at least $\tau_{\min}=0.4$\,s. Given a matching tolerance $\delta$, a source pause and a hypothesis pause may be matched one-to-one when the difference between their center times is at most $\delta$. Matched pairs are true positives, unmatched source pauses are false negatives, and unmatched hypothesis pauses are false positives:
\begin{align}
    P
    &=
    \frac{\mathrm{TP}}{\mathrm{TP}+\mathrm{FP}},
    &
    R
    &=
    \frac{\mathrm{TP}}{\mathrm{TP}+\mathrm{FN}},
    &
    \mathrm{F1}
    &=
    \frac{2PR}{P+R}.
\end{align}
We report F1@0.4 using $\delta=0.4$\,s. Higher is better.

\paragraph{Pause-span tIoU.}
Temporal intersection over union measures the overlap between the source and hypothesis silent regions:
\begin{equation}
    \mathrm{tIoU}(S,H)
    =
    \frac{
        \mu(S\cap H)
    }{
        \mu(S\cup H)
    }.
\end{equation}
It jointly penalizes missed pauses, additional pauses, temporal shifts, and mismatched pause durations. Higher is better.

\paragraph{Structure Error Rate (SER).}
SER decomposes pause-timeline disagreement into three components:
\begin{align}
    E_{\mathrm{miss}}
    &=
    \mu(S\setminus H),
    &
    E_{\mathrm{fa}}
    &=
    \mu(H\setminus S),
    &
    E_{\mathrm{dur}}
    &=
    \left|
        T^{\mathrm{hyp}}-T^{\mathrm{src}}
    \right|.
\end{align}
Here, $E_{\mathrm{miss}}$ is the duration for which the source is silent but the translation is speaking, $E_{\mathrm{fa}}$ is the duration for which the translation is silent but the source is speaking, and $E_{\mathrm{dur}}$ measures utterance-end mismatch. We normalize their sum by the source duration:
\begin{equation}
    \mathrm{SER}
    =
    \frac{
        E_{\mathrm{miss}}
        +
        E_{\mathrm{fa}}
        +
        E_{\mathrm{dur}}
    }{
        T^{\mathrm{src}}
    }.
\end{equation}
Lower is better. Unlike tIoU, SER also exposes the contributions of the different error types.

\paragraph{Pause-onset error.}
Let $\mathcal{M}_{\delta}$ be the pause pairs matched under tolerance $\delta$, and let $b(p)$ denote the onset time of pause $p$. We compute the median absolute onset error over the matched pairs:
\begin{equation}
    \mathrm{onsetMedAE}@\delta
    =
    \underset{(p,q)\in\mathcal{M}_{\delta}}{\operatorname{median}}
    \left|
        b(p)-b(q)
    \right|.
\end{equation}
We report onsetMedAE@0.4 in seconds using the same matches as F1@0.4, and abbreviate it as onMedAE in the tables. Lower is better. Because this metric is conditioned on successfully matched pauses, it must be interpreted jointly with pause F1.

\section{Human Evaluation}
\label{app:human_eval}

\paragraph{Participants and setup.}
Participants were recruited through Prolific and were adults aged 22--58 years residing in the United Kingdom and United States. A total of 25 participants took part in the study. They were required to self-confirm fluency in English and Mandarin Chinese before participation. System identities were concealed, and output presentation order was randomized within each source utterance. Each output was evaluated on two separate five-point scales.

\paragraph{Audio naturalness (Nat.).}
Participants assessed how natural the Mandarin speech sounded, considering pronunciation, rhythm, phrasing, intonation, fluency, pauses, and noticeable speech artifacts. They judged this dimension independently of translation accuracy and similarity to the original speaker. Scores ranged from 1 (\emph{very unnatural}) to 5 (\emph{completely natural}).

\paragraph{Translation quality (Qual.).}
Participants assessed how accurately and completely the Mandarin speech conveyed the English source's meaning and important information. Natural rephrasing was permitted provided that the original meaning was preserved. Scores ranged from 1 (\emph{meaning not conveyed}) to 5 (\emph{fully accurate and complete}).

\begin{figure}[htb]
    \centering
    \includegraphics[
        width=0.95\linewidth,
        trim={1cm 102cm 25cm 6cm}, clip
    ]{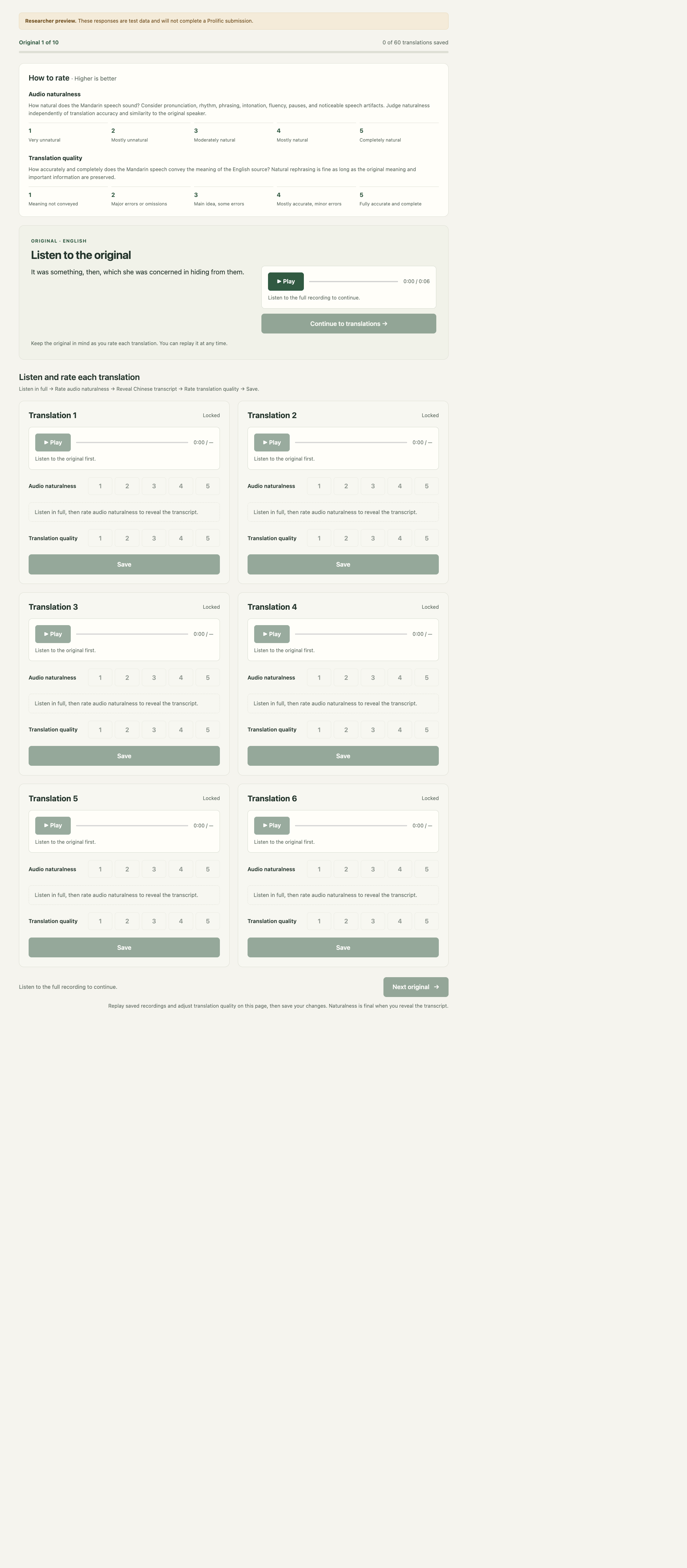}
    \caption{Human evaluation interface for audio naturalness and translation quality. Participants rate naturalness before viewing the Mandarin transcript, which is subsequently revealed for translation-quality assessment.}
    \label{fig:human_eval}
\end{figure}

\paragraph{Rating procedure.}
As illustrated in Fig.~\ref{fig:human_eval}, participants first listened to the English source with its transcript. They then listened to each Mandarin output in full and rated audio naturalness before its transcript was revealed. Subsequently, they assessed translation quality with the Mandarin transcript available. Whenever the transcript differed from the spoken output, participants were instructed to judge the speech rather than the displayed text.

\section{Additional Experimental Results and Ablations}
\label{app:additional-experiments}

This appendix complements Section~\ref{sec:experiments} with analyses of decoding sensitivity, domain-specific translation quality, Stage~3 supervision composition, and alternative approaches to temporal control. 

\begin{figure}[htb]
    \centering
    \begin{subfigure}{0.48\linewidth}
        \centering
        \includegraphics[width=\linewidth]{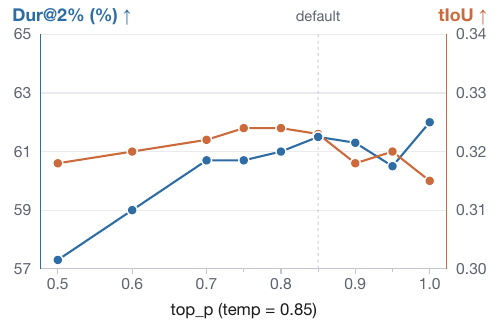}
        \caption{Top-$p$ sweep}
        \label{fig:decode-sweep-top-p}
    \end{subfigure}
    \hfill
    \begin{subfigure}{0.48\linewidth}
        \centering
        \includegraphics[width=\linewidth]{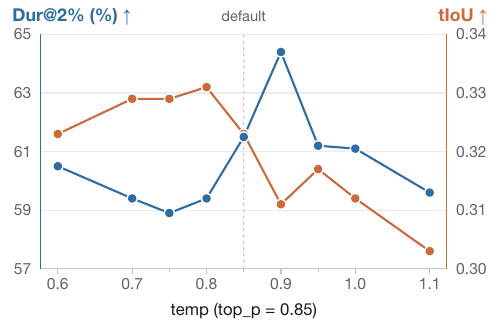}
        \caption{Temperature sweep}
        \label{fig:decode-sweep-temp}
    \end{subfigure}
    \caption{Top-p and temperature decoding settings. Left axes show duration accuracy (Dur@2\%); right axes show pause overlap (tIoU). Dashed lines mark the default settings.}
    \label{fig:decode-sweep}
\end{figure}

\paragraph{Sensitivity to decoding parameters.}
Figure~\ref{fig:decode-sweep} varies the nucleus-sampling threshold $p$ and temperature separately, keeping the other parameter fixed at its default value of 0.85. The two temporal metrics do not improve monotonically together. In the top-$p$ sweep, restricting sampling to $p=0.5$ reduces strict duration agreement, while intermediate values provide better pause overlap than $p=1.0$. The temperature sweep exhibits a clearer trade-off: increasing the temperature from 0.85 to 0.9 improves $\durat{2\%}$ but reduces tIoU, and further increasing it to 1.1 degrades both metrics relative to the default. Thus, the default setting of $p=0.85$ and temperature 0.85 provides a compromise between endpoint fidelity and local pause alignment rather than maximizing either metric in isolation.

\paragraph{Domain-specific translation quality.}
Table~\ref{tab:domain-quality} reports translation quality separately for CoVoST2 and GigaSpeech on the ASR-mediated track. On CoVoST2, \model achieves 0.745 COMET and 5.18 MetricX-QE, compared with 0.738 and 5.68 for \uniss. The difference is larger on GigaSpeech, where COMET increases from 0.732 to 0.809 and MetricX-QE decreases from 5.64 to 3.61. The reported advantage therefore appears in both the predominantly read-speech and spontaneous-speech subsets, although its magnitude differs substantially. These are within-domain comparisons; differences in absolute scores across the two subsets should not be attributed solely to speaking style, since their reference construction also differs, as described in Section~\ref{sec:setup}.

\begin{table}[htbp]
  \centering
    \caption{Domain-specific translation quality on the ASR-mediated track. Unlike Table~\ref{tab:main}, these scores are computed from ASR transcriptions of the generated speech and therefore also reflect speech realization errors.}  
    \label{tab:domain-quality}
  \small
  \begin{tabular}{llcc}
    \toprule
    Domain & System & COMET $\uparrow$ & MetricX-QE $\downarrow$ \\
    \midrule
    \multirow{2}{*}{CoVoST2} & \uniss & 0.738 & 5.68 \\
      & \model & \textbf{0.745} & \textbf{5.18}\\
    \midrule
    \multirow{2}{*}{GigaSpeech} & \uniss & 0.732 & 5.64 \\
      & \model & \textbf{0.809} & \textbf{3.61}\\
    \bottomrule
  \end{tabular}
\end{table}

\paragraph{Composition of timeline-preserving supervision.}
Table~\ref{tab:stage3_k0_ablation} compares Stage~3 training with only no-pause examples, only pause-aligned examples, and their union. No-pause supervision yields the strongest translation scores in this comparison, with 0.793 COMET and 3.77 MetricX-QE, but provides limited local alignment: pause F1 and tIoU reach only 0.180 and 0.134. Using only pause-aligned examples raises pause F1 to 0.523, but achieves lower strict duration agreement, with $\durat{2\%}$ of 28.1\% compared with 33.4\% for no-pause supervision. Combining both subsets yields the best $\durat{2\%}$ and $\durat{5\%}$, at 48.6\% and 83.3\%, together with the highest tIoU of 0.293 and lowest SER of 0.198. Its pause F1 of 0.518 remains close to that of pause-aligned-only training, although its translation scores do not match those of no-pause-only training. Combining the two subsets improves duration accuracy and pause-span overlap. No-pause-only training gives better translation scores, while pause-aligned-only training gives a slightly higher pause F1.

\begin{table*}[htbp]
\centering
\caption{Ablation of the Stage-3 supervision composition.}
\label{tab:stage3_k0_ablation}
\scriptsize
\setlength{\tabcolsep}{4.0pt}
\begin{tabular}{lccccccc}
\toprule
Stage-3 supervision
& COMET $\uparrow$
& MetricX-QE $\downarrow$
& Dur@2\% $\uparrow$
& Dur@5\% $\uparrow$
& Pause F1 $\uparrow$
& tIoU $\uparrow$
& SER $\downarrow$ \\
\midrule

$\mathcal{D}_{no{\text -}pause}$
& \textbf{0.793} & \textbf{3.77} & 33.4 & 63.9 & 0.180 & 0.134 & 0.258 \\

$\mathcal{D}_{\mathrm{pause{\text -}aligned}}$
& 0.767 & 4.52 & 28.1 & 59.7 & \textbf{0.523} & 0.284 & 0.230 \\

Full ($\mathcal{D}_{no{\text -}pause} \cup \mathcal{D}_{\mathrm{pause{\text -}aligned}}$)
& 0.772 & 4.38 & \textbf{48.6} & \textbf{83.3} & 0.518 & \textbf{0.293} & \textbf{0.198} \\

\bottomrule
\end{tabular}
\end{table*}

\paragraph{Initialization for policy optimization.}
Table~\ref{tab:rl-init} examines RL initialization from different training stages. The Stage~1 checkpoint yields no valid rollouts, while Stage~2 achieves a 93.8\% valid-rollout rate. However, RL from Stage~2 leaves $\durat{2\%}$ unchanged at 5.2\% and improves tIoU only from 0.086 to 0.097. In contrast, Stage~3 provides a 100\% valid-rollout rate and enables RL to raise $\durat{2\%}$ from 48.6\% to 61.5\% and tIoU from 0.293 to 0.322. Thus, valid speech generation alone is insufficient for substantial temporal gains in this setting. RL is more effective at refining timeline-following behavior established through supervised training than at acquiring it from natural-pace speech generation alone.

\begin{table*}[htbp]
  \centering
  \caption{Reinforcement learning initialized from text grounding (Stage 1), natural-pace speech training (Stage 2), or timeline-preserving supervision (Stage 3).}
  \label{tab:rl-init}
  \footnotesize
  \setlength{\tabcolsep}{5pt}
  \renewcommand{\arraystretch}{1.1}

  \begin{tabular*}{\textwidth}{@{\extracolsep{\fill}}c cc cc cc c@{}}
    \toprule
    \multirow{3}{*}{RL initialization}
    & \multicolumn{2}{c}{COMET $\uparrow$}
    & \multicolumn{2}{c}{$\durat{2\%}$ (\%) $\uparrow$}
    & \multicolumn{2}{c}{tIoU $\uparrow$}
    & Valid rollout rate \\
    \cmidrule(lr){2-3}
    \cmidrule(lr){4-5}
    \cmidrule(lr){6-7}
    & Pre-RL & Post-RL & Pre-RL & Post-RL & Pre-RL & Post-RL & \\
    \midrule
    Stage~1 checkpoint & 0.823 & N/A             & --   & N/A           & --    & N/A            & 0\% \\
    Stage~2 checkpoint & 0.823 & \textbf{0.829} & 5.2  & 5.2           & 0.086 & 0.097          & 93.8\% \\
    Stage~3 checkpoint & 0.772 & 0.788           & 48.6 & \textbf{61.5} & 0.293 & \textbf{0.322} & 100\% \\
    \bottomrule
  \end{tabular*}
\end{table*}

\paragraph{Global duration correction versus local timeline preservation.}
Table~\ref{tab:global_stretch} examines whether global time-scale stretching after generation can recover source-timeline alignment. All listed systems reach 100\% $\durat{2\%}$ after stretching, but substantial differences in local alignment remain. The stretched baselines obtain pause F1 scores of 0.026--0.159 and tIoU scores of 0.034--0.116, still well below the 0.550 pause F1 and 0.322 tIoU achieved by \model without stretching. For \uniss, correcting global duration raises pause F1 only from 0.118 to 0.131. Stretching \model produces comparatively small additional gains, increasing pause F1 from 0.550 to 0.561 and tIoU from 0.322 to 0.334. The reductions in SER must also be interpreted in light of its definition: correcting duration directly reduces its endpoint-mismatch component and therefore does not, by itself, establish accurate internal pause placement. Uniformly rescaling an utterance can compensate for a global rate mismatch, but cannot independently adjust individual speaking regions and pauses. This experiment supports the distinction between matching the utterance endpoint and preserving its internal timeline. Stretching is used only for this diagnostic comparison; the main \model results do not use post-hoc temporal correction.

\begin{table*}[htb]
  \centering
  \caption{Impact of global time-scale stretching on temporal alignment. Each entry reports the performance before and after stretching (\textit{before} $\rightarrow$ \textit{after}).}
  \label{tab:global_stretch}
  \small
  \begin{adjustbox}{max width=\textwidth}
  \begin{tabular}{lccccc}
    \toprule
    System
    & $\durat{2\%}$ $\uparrow$
    & Pause F1 $\uparrow$
    & tIoU $\uparrow$
    & SER $\downarrow$
    & onMedAE $\downarrow$ \\
    \midrule

    Seamless-L
    & 3.7 $\rightarrow$ 100
    & 0.002 $\rightarrow$ 0.026
    & 0.008 $\rightarrow$ 0.034
    & 0.407 $\rightarrow$ 0.143
    & 0.533 $\rightarrow$ 0.252 \\

    Seamless-Ex
    & 4.0 $\rightarrow$ 100
    & 0.062 $\rightarrow$ 0.124
    & 0.059 $\rightarrow$ 0.101
    & 0.395 $\rightarrow$ 0.177
    & 0.230 $\rightarrow$ 0.243 \\

    Qwen2.5-Omni-7B
    & 4.0 $\rightarrow$ 100
    & 0.080 $\rightarrow$ 0.159
    & 0.060 $\rightarrow$ 0.114
    & 0.446 $\rightarrow$ 0.177
    & 0.211 $\rightarrow$ 0.227 \\

    \uniss
    & 25.6 $\rightarrow$ 100
    & 0.118 $\rightarrow$ 0.131
    & 0.108 $\rightarrow$ 0.116
    & 0.288 $\rightarrow$ 0.201
    & 0.266 $\rightarrow$ 0.262 \\

    \midrule

    \model
    & 61.5 $\rightarrow$ 100
    & 0.550 $\rightarrow$ \textbf{0.561}
    & 0.322 $\rightarrow$ \textbf{0.334}
    & 0.128 $\rightarrow$ \textbf{0.103}
    & 0.162 $\rightarrow$ \textbf{0.156} \\

    \bottomrule
  \end{tabular}
  \end{adjustbox}
\end{table*}

\begin{table*}[htbp]
\centering
\caption{Comparison with F5-TTS resynthesis of \model translations and the Stage~3 construction pipeline.}
\label{tab:construction_pipeline_f5_ablation}

\small
\setlength{\tabcolsep}{3.5pt}
\resizebox{\textwidth}{!}{%
\begin{tabular}{lccccccccc}
\toprule
System
& Success (\%) $\uparrow$
& COMET $\uparrow$
& MetricX-QE $\downarrow$
& CER (\%) $\downarrow$
& $\durat{10\%}\uparrow$
& Pause F1 $\uparrow$
& tIoU $\uparrow$
& SER $\downarrow$
& onMedAE $\downarrow$ \\
\midrule

\model
& 99.4 & 0.788 & 4.12 & 2.68 & 98.7 & 0.550 & 0.322 & 0.128 & 0.162 \\

\model translated text + F5-TTS
& 100.0 & 0.788 & 4.12 & 8.14 & 100.0 & 0.263 & 0.172 & 0.216 & 0.277 \\

Stage-3 construction pipeline
& 47.5 & 0.724 & 4.88 & 5.07 & 100.0 & \textbf{0.642} & \textbf{0.481} & \textbf{0.081} & \textbf{0.067} \\

\bottomrule
\end{tabular}%
}
\end{table*}

\paragraph{Comparison with synthesis-based alternatives.}
Table~\ref{tab:construction_pipeline_f5_ablation} compares \model with F5-TTS~\citep{chen2025f5} resynthesis of its translated text and the Stage~3 construction pipeline. Under the evaluated configuration, F5-TTS achieves 100\% $\durat{10\%}$ but yields lower pause F1 (0.263 vs.\ 0.550) and higher CER (8.14\% vs.\ 2.68\%), despite using the same translated text. The Stage~3 pipeline obtains better pause-alignment scores but succeeds on only 47.5\% of inputs, compared with 99.4\% for \model; its temporal scores should therefore be interpreted alongside this substantially lower output coverage.

\section{Limitations}
Our experiments focus on English-to-Chinese translation of short utterances, leaving generalization to other language pairs and long-form speech unverified. Moreover, our timeline-preserving supervision relies on automatically translated and synthesized targets. Despite quality filtering, these targets may inherit errors or biases from the construction pipeline, while rejecting utterances that cannot satisfy local duration constraints may favor examples that are easier to synchronize.

Source-timeline preservation is an application-specific objective that may constrain translation flexibility and natural target-language phrasing. Our formulation captures only a subset of temporal structure, focusing on utterance-level duration and selected speech–silence events, while not explicitly modeling finer-grained prosody, word-level timing, semantic alignment, or audiovisual lip synchronization. Temporal control also remains imperfect.

\end{document}